\documentclass[]{fairmeta}
\usepackage[dvipsnames]{xcolor}
\usepackage[most]{tcolorbox}

\usepackage[suppress]{color-edits}
\addauthor{john}{cyan}
\addauthor{ofir}{magenta}
\addauthor{kn}{blue}
\addauthor{emily}{teal}
\addauthor{kilian}{BurntOrange}
\addauthor{jeff}{brown}
\addauthor{sten}{Red}
\addauthor{parth}{Turquoise}

\usepackage{caption}
\usepackage{csquotes}
\usepackage{longtable}
\usepackage[bottom]{footmisc}

\title{ProgramBench: Can Language Models Rebuild Programs From Scratch?}

\author[1,3*]{John Yang}
\author[1*]{Kilian Lieret}
\author[1,4]{Jeffrey Ma}
\author[1]{Parth Thakkar}
\author[1]{Dmitrii Pedchenko}
\author[1]{Sten Sootla}
\author[1]{Emily McMilin}
\author[2]{Pengcheng Yin}
\author[2]{Rui Hou}
\author[1]{Gabriel Synnaeve}
\author[3]{Diyi Yang}
\author[1]{Ofir Press}

\affiliation[1]{Meta FAIR}
\affiliation[2]{Meta TBD}
\affiliation[3]{Stanford University}
\affiliation[4]{Harvard University}
\contribution[*]{Equal Contribution}

\abstract{
Turning ideas into full software projects from scratch has become a popular use case for language models.
Agents are being deployed to seed, maintain, and grow codebases over extended periods with minimal human oversight.
Such settings require models to make high-level software architecture decisions.
However, existing benchmarks measure focused, limited tasks such as fixing a single bug or developing a single, specified feature.
We therefore introduce \bench{} to measure the ability of software engineering agents to develop software holisitically.
In \bench{}, given only a program and its documentation, agents must architect and implement a codebase that matches the reference executable's behavior.
End-to-end behavioral tests are generated via agent-driven fuzzing, enabling evaluation without prescribing implementation structure.
Our 200 tasks range from compact CLI tools to widely used software such as \texttt{FFmpeg}, \texttt{SQLite}, and the \texttt{PHP} interpreter.
We evaluate 9 LMs and find that none fully resolve any task, with the best model passing 95\% of tests on only 3\% of tasks. Models favor monolithic, single-file implementations that diverge sharply from human-written code.
}

\date{\today}
\correspondence{John Yang at \email{johnby@meta.com}, Kilian Lieret at \email{klieret@meta.com}}

\metadata[Code]{\url{https://github.com/facebookresearch/ProgramBench}}

\newcommand{\bench}{ProgramBench}
\newcommand{\smallperc}{{\footnotesize\%}}

\definecolor{PromptBlue}{HTML}{1E5AA8}
\definecolor{PromptRed}{HTML}{C0392B}
\definecolor{PromptGreen}{HTML}{27AE60}

\newtcblisting{verbpromptboxenv}[1][PromptBlue]{
  colframe=#1,
  colback=#1!8,
  boxrule=0.8pt,
  arc=2pt,
  left=4pt,right=4pt,top=2pt,bottom=2pt,
  breakable,
  before skip=0.5\baselineskip,
  after skip=0.5\baselineskip,
  listing only,
  listing options={
    basicstyle=\ttfamily\footnotesize,
    breaklines=true,
    breakatwhitespace=false,
    columns=fullflexible,
    keepspaces=true,
    aboveskip=0pt,
    belowskip=0pt,
  }
}
\newcommand{\codebox}[2][PromptBlue]{%
  \tcbinputlisting{
    colframe=#1,
    colback=#1!8,
    coltext=black,
    boxrule=0.8pt,
    arc=2pt,
    left=18pt,right=4pt,top=2pt,bottom=2pt,
    breakable,
    before skip=0.5\baselineskip,
    after skip=0.5\baselineskip,
    listing only,
    listing file={#2},
    listing options={
      basicstyle=\ttfamily\footnotesize\color{black},
      numbers=left,
      numberstyle=\tiny\color{gray},
      numbersep=6pt,
      breaklines=true,
      breakatwhitespace=false,
      columns=fullflexible,
      keepspaces=true,
      aboveskip=0pt,
      belowskip=0pt,
    }
  }%
}
\newcommand{\verbpromptboxfile}[2][PromptBlue]{%
  \tcbinputlisting{
    colframe=#1,
    colback=#1!8,
    coltext=black,
    boxrule=0.8pt,
    arc=2pt,
    left=4pt,right=4pt,top=2pt,bottom=2pt,
    breakable,
    before skip=0.5\baselineskip,
    after skip=0.5\baselineskip,
    listing only,
    listing file={#2},
    listing options={
      basicstyle=\ttfamily\footnotesize\color{black},
      breaklines=true,
      breakatwhitespace=false,
      columns=fullflexible,
      keepspaces=true,
      aboveskip=0pt,
      belowskip=0pt,
    }
  }%
}

\begin{document}

\maketitle

\section{Introduction}
\label{sec:introduction}

\begin{figure}[b]
\centering
\includegraphics[width=\textwidth]{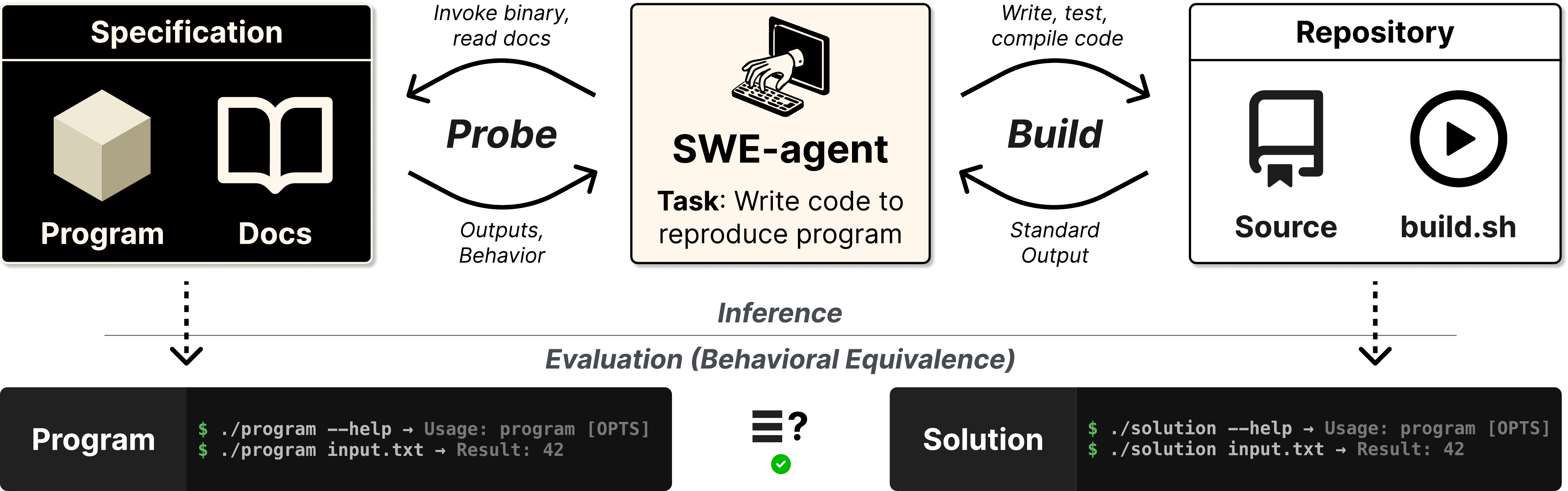}
\caption{
\textbf{\bench{}} evaluates models on their ability to write software projects from scratch.
Given a software program (e.g., executable) and its documentation, a software engineering agent (SWE-agent) is tasked with producing source code and a build script that reconstructs the original program's behavior.
}
\label{fig:preview}
\end{figure}

Language Models (LMs) are increasingly being used to turn ideas expressed in natural language into full-fledged code repositories~\citep{carlini2026building,cursor_scaling_agents,replit}.
Unlike smaller scope tasks such as function generation~\citep{hendrycks2021measuringcodingchallengecompetence} or GitHub issue resolution~\citep{jimenez2024swebenchlanguagemodelsresolve}, which typically demand understanding a pre-existing codebase well enough to make localized changes, building a functional application from scratch requires models to engage heavily with software design~\citep{jansen2005software}.

To understand what this entails, consider how a human programmer approaches the same task.
Before a single line is written, she asks herself a series of important questions:
What programming language and build system should be used?
How should the codebase be organized?
What data structures should represent the program's core entities?
How should errors be detected and communicated?
Such requisite questions, which developers constantly revisit throughout the development lifecycle, lead to pivotal design decisions that shape the codebase far more profoundly than any individual code change.
Although we are progressively entrusting LMs to similarly build software from the ground up, the ability of LMs to make such architectural decisions, choose abstractions, and decompose a system into coherent modules has not been studied extensively.

To bridge this gap, we introduce \bench{}, a benchmark that challenges software engineering (SWE) agents to produce code that recovers the functionality of a software program (e.g., executables, \texttt{.dmg}'s, \texttt{.pkg}'s).
Given a program and documentation, a SWE-agent, defined as an LM equipped with an agent scaffold to interact with a terminal environment~\citep{yang2024sweagentagentcomputerinterfacesenable}, must write source code and a compile script that reproduces the original program's behavior.
Every software design decision is entirely the model's to make.

We synthesize \bench{} tasks from open-source GitHub repositories.
First, we identify repositories written in compiled languages (e.g., C/C++, Golang, Rust, Java) that build a program.
Next, to convert a repository into a task instance, we compile the program, then strip away all source code and tests, leaving only the program and its documentation as the task's starting point.

To evaluate a model's solution, we generate behavioral tests by prompting a SWE-agent to systematically probe the original program with varied inputs and codify the observed input-output behavior into assertions that a candidate reconstruction must satisfy.
Crucially, these tests are never revealed to the task worker.
Since tests target executable behavior rather than source code, evaluation is entirely implementation agnostic; a model may use different algorithms, abstractions, or even programming languages than the original codebase, and still pass as long as the input-output behavior matches.
While any test suite necessarily under-approximates an executable's full specification, we empirically demonstrate that our test generation pipeline creates large suites that reliably capture core functionality.

Using our pipeline, we collect 200 task instances, ranging from compact CLI tools to complex, widely used software including language interpreters (\texttt{PHP}, \texttt{Lua}, \texttt{tinycc}), databases (\texttt{DuckDB}, \texttt{SQLite}), media and compression utilities (\texttt{FFmpeg}, \texttt{zstd}, \texttt{xz}), and developer tools (\texttt{ripgrep}, \texttt{fzf}, \texttt{jq}).
We evaluate 9 language models equipped with \texttt{mini-SWE-agent}, a widely adopted coding agent scaffold for open source SWE-agent research.
The results resoundingly confirm \bench{}'s difficulty for today's models; no task instance is fully resolved. 
However, test pass rates are significantly different between models.
The best model, Opus 4.7, manages to pass 95\% of tests for 3\% of task instances.
Further analysis reveals that model-written codebases diverge significantly from human-written ones, favoring monolithic file structures with longer functions.
Our trajectory analyses showcase how models vary in the length and make up of the way they develop software.

We open source \bench{} to enable the community to reproduce and build upon our investigations.

\section{\bench}
\label{sec:benchmark}

This section describes \bench{} in detail.
We first formalize the task (\S\ref{sec:benchmark:formulation}), then explain how task instances are semi-automatically constructed from open-source repositories (\S\ref{sec:benchmark:construction}).
Finally, we review benchmark statistics (\S\ref{sec:benchmark:statistics}) and distinguishing features (\S\ref{sec:benchmark:features}).

\subsection{Task Formulation}
\label{sec:benchmark:formulation}
Given a gold (reference) executable and its usage documentation, a task worker is asked to write source code and a build script that constructs a candidate executable which should reproduce the behavior of the gold executable.
The task worker does not have access to the internet and is free to implement the solution in any programming language.
The task worker is informed of these conditions via the initial prompt, and no-internet is enforced by running the Docker container without internet.

To evaluate, we run a generated test suite, where each test checks whether the candidate executable exhibits the same observable behavior as the gold executable for a given input (e.g., matching standard output, exit codes, or file system side effects).
The test suite is never revealed at any point to the task worker.
While any test suite checks a finite set of inputs and therefore necessarily under-approximates the gold executable's full specification, \bench{}'s framing makes extending test coverage trivial.

\begin{figure}[t]
\centering
\includegraphics[width=\linewidth]{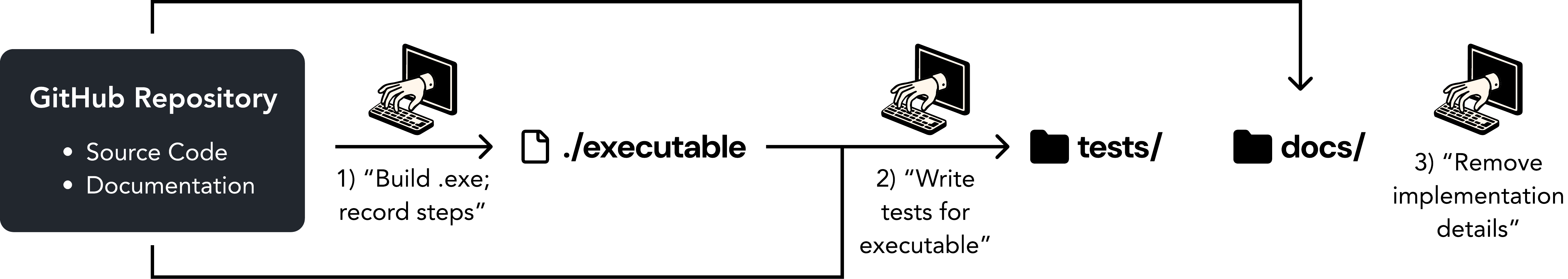}
\caption{
\textbf{\bench{} task collection pipeline.}
To turn a GitHub repository into an \bench{} task, we use a SWE-agent to compile an executable, generate behavioral tests, and strip away implementation details.
The sourcing workflow only requires a repository to produce an executable or program, making it extensible to many codebases.}
\label{fig:collection_pipeline}
\end{figure}

\subsection{Benchmark Construction}
\label{sec:benchmark:construction}
Next, we discuss our four-stage pipeline for converting open source GitHub repositories into \bench{} task instances, as visualized in Figure~\ref{fig:collection_pipeline}.
All construction steps use the \texttt{mini-SWE-agent}\footnote{\url{https://mini-swe-agent.com}} harness with Claude Sonnet~4.5, operating inside a Docker container based on \texttt{ubuntu:22.04}.

\textbf{Identify candidate repositories.} We synthesize \bench{} tasks from open source GitHub repositories.
First, we filter for repositories that may produce a standalone executable or program.
A strong heuristic is to look for projects written in compiled languages (e.g., C/C++, Golang, Rust).

\textbf{Construct executable from source.} Given a repository, we task a SWE-agent with compiling the gold executable and, if successful, record the commands that reproduce the build in a single build script (Step 1 in Figure~\ref{fig:collection_pipeline}).

\textbf{Generate behavioral tests.} We use an agent to explore the program, its source code, existing tests, and documentation, and then generate behavioral tests (Step 2 of Figure~\ref{fig:collection_pipeline}).
The test assertions target externally observable effects rather than source-level internals.
For example, a test might assert that specific strings appear in stdout or stderr, or that an invocation produces expected files.
The agent is also prompted to identify and include in its test suite any existing behavioral tests defined in the repository (\textit{harvesting}).

The agent continuously measures the line coverage of the current test suite and iteratively writes new tests to invoke missing code paths, attempting to achieve full coverage.

Some tests may have missing or trivially true assertions.
Therefore, to ensure assertion quality, tests are flagged if they fail the gold binary or trigger our assertion quality linter (Appendix~\ref{app:lint-rules}), which detects structurally weak assertion patterns such as exit-code-only checks, short substring matches, and disjunctive assertions.
The agent is prompted to revise all flagged tests.
At the end, any remaining tests that do not pass with the gold binary deterministically or pass a dummy binary are discarded.
More details in \S\ref{appx:benchmark:test_generation}.



Prior work has shown that coding benchmarks, such as SWE-bench, have some test suites with one of the following two shortcomings~\citep{chowdhury2024introducing}.
First, a task's test suite could be overly stringent, meaning it checks for criteria not apparent from the initial task definition.
Second, a task's test suite might not fully check whether a solution actually solves the initially stated task.
For the first case, \bench{} tests assert only on observable behavior of the reference executable, precluding overspecification of source-level internals.
A subtler concern is whether tests demand exact reproduction of implementation-dependent output (e.g., floating point precision or rendering discretization).
We address this directly: the model has full access to the gold executable at inference time, so any behavior a behavioral test expects is discoverable by running that same command.
An audit of all 200 task instances found zero tests that invoke flags or subcommands not surfaced by the executable's documentation, and only 5 instances where implementation-dependent output could plausibly appear, none of which contained such assertions in practice (\S\ref{appx:benchmark:test_generation:over_specification}).
For the second case, a finite test suite necessarily under-approximates an executable's full specification \citep{liu2023codegeneratedchatgptreally, legoues2015manyplus}.
We quantify this risk in \S\ref{sec:analysis}: our generated suites achieve line coverage broadly comparable to the native test suites shipped by developers in the same repositories.


\textbf{Build an inference environment.} The starting state for each task consists of the gold executable and usage-related documentation.
To construct the Docker image for the starting state, we first obtain usage-related documentation removing source code and any implementation details from the repository with a SWE-agent (Step 3 in Figure~\ref{fig:collection_pipeline}).
The compiled executable is then injected into an independent docker image as the only artifact carried over from the build step (Step 1).
The reason for copying the executable over, rather than simply rebuilding from source code, is to ensure there are no local build artifacts or dependency caches that could reveal the original program's implementation.
The executable is also set to execute-only permissions to prevent reading or reverse engineering of the binary with a tool like \texttt{ghidra}\footnote{\url{https://github.com/nationalsecurityagency/ghidra}}.
Lastly, we also include test assets that a model cannot reasonably synthesize on its own (e.g., images, domain-specific binary formats).
We review inference guidelines thoroughly in \S\ref{appx:benchmark:inference}.

\subsection{Dataset Statistics}
\label{sec:benchmark:statistics}
\begin{figure}[t]
    \begin{minipage}[b]{0.38\textwidth}
        \centering
        \includegraphics[width=0.85\textwidth]{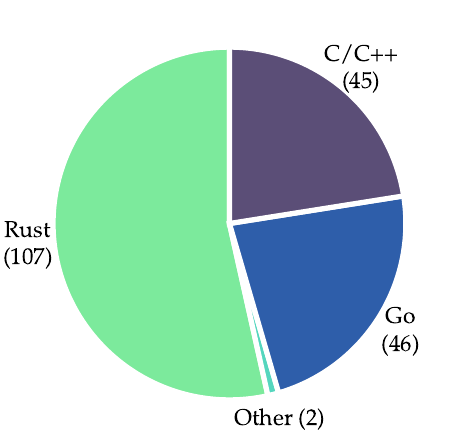}
        \captionof{figure}{
            Distribution of programming languages across \bench{} task instances.
            To solve the task, models may write their solution in any language they choose.}
        \label{fig:dist_language}
    \end{minipage}
    \hfill
    \begin{minipage}[b]{0.58\textwidth}
        \centering
        \centering
\small
\begin{tabular}{@{}lrrr@{}}
\toprule
\textbf{Metric} & \textbf{Median} & \textbf{Min} & \textbf{Max} \\
\midrule
Code lines & 8,635 & 212 & 2,701,283 \\
Code files & 50 & 1 & 5,342 \\
Runtime dependencies & 10 & 0 & 113 \\
Max directory depth & 3 & 0 & 13 \\
\midrule
Tests & 770 & 224 & 14,645 \\
\midrule
GitHub stars & 2,124 & 202 & 79,693 \\
Contributors & 22 & 1 & 422 \\
Commits & 646 & 13 & 145,991 \\
Repo age (years) & 7.9 & 0.3 & 17.8 \\
\bottomrule
\end{tabular}
\captionof{table}{Summary statistics for the 200 \bench{} task instances, illustrating the diversity of the dataset across codebase scale, dependency complexity, and development history. Top rows cover codebase statistics, while bottom rows quantify community contribution.}
\label{tab:summary_stats}

    \end{minipage}
\end{figure}
We created 200 task instances from open-source repositories spanning compression, language interpreters, visualization, linting, text processing, and more (Figure~\ref{fig:dist_language}, Table~\ref{tab:summary_stats}).
Tasks range from small CLI tools to large-scale projects such as FFmpeg and the PHP interpreter, and are mostly written in Rust, Go, or C/C++, with one project in Java and one in Haskell.
Our evaluation suite totals 248,853 test functions across all instances (median 770 per task); additional test generation analyses and dataset statistics in \S\ref{appx:benchmark:test_generation} and \S\ref{appx:benchmark:statistics}.

\subsection{Task Features}
\label{sec:benchmark:features}
\textbf{Open-ended software design.}
In \bench{}, models receive only an executable and documentation.
Every architectural decision, from choice of language to module decomposition to data structure design, is the model's to make.
There is no skeleton, mandated abstractions, or preset file organization.
Because evaluation compares executable behavior rather than source code, \bench{} admits many valid solutions; a model may choose entirely different languages, algorithms, or architecture and still pass, making models' design choices directly comparable across the same task.
This is what makes \bench{} a test of software design, not just implementation alone.

\textbf{Burden to discover specifications.}
In \bench{}, the executable serves as a comprehensive but opaque oracle.
Expected behavior is fully encoded, but must be queried to be understood.
Importantly, the model is not probing blindly: it has access to the program's documentation and help output, which surface available flags and subcommands (\S\ref{sec:benchmark:construction}).
In practice, interacting with the executable resembles how a developer queries a product manager: ``when a user runs \texttt{X} with flag \texttt{-y}, what should the output be?''
The model must decide which questions to ask and in what order.
This setting tests a model's ability to infer behavior through systematic, hypothesis-driven exploration, mirroring challenges developers routinely face when probing partially documented APIs or onboarding to unfamiliar systems by observing behavior in the absence of complete specifications.

\textbf{Simple collection criteria.}
Our collection pipeline requires only that a repository produce a standalone executable.
No existing test suite, language-specific AST tooling, or ecosystem-reliant test frameworks are needed.
The benchmark is therefore straightforward to extend with new instances over time, a valuable property for sustaining benchmark relevance~\citep{deng2025swebenchproaiagents,zhang2025swebenchgoeslive}.
The same pipeline can also generate training data, similar to how prior benchmark collection schemas have been repurposed for training~\citep{badertdinov2025swe,pham2025swe,yang2025swesmithscalingdatasoftware}.

\section{Experiments}
\label{sec:experiments}

\textbf{Models.} We evaluate $9$ recent language models that are regarded as strong coding models based on rank on existing benchmarks, including SWE-bench and Terminal-bench.
These include Claude Opus~4.7, Claude Opus~4.6, Claude Sonnet~4.6, Claude Haiku~4.5, Gemini~3.1 Pro, Gemini~3 Flash, GPT~5.4, GPT~5.4 mini, and GPT~5 mini. 
All models use vendor-default hyperparameters.

\textbf{Agent scaffold.}
We use \texttt{mini-SWE-agent} because it is both widely adopted as a baseline by other benchmarks (SWE-bench Verified, Multilingual\footnotemark[\value{footnote}], Terminal-bench\footnote{\url{https://tbench.ai}}), and deliberately minimal in its scaffolding, reducing confounds between model capability and harness design.
Each model runs inside a container with 20 CPUs and 60GB RAM, with a limit of 1,000 steps and 6 hours per run; full configuration details are in \S\ref{appx:results:exp_setup}.

\textbf{Metrics.} We primarily report \textbf{\% Resolved}, which refers to the percentage of task instances where the model's codebase passes \textit{all} associated test cases and is not flagged as cheating.
Due to the challenging nature of this benchmark, we also report \textbf{\% Tests Passed} per task instance, which captures partial progress even when no task is fully resolved.
However, we note that this softer metric is only meaningful for relative comparisons between models; even a single failed test can imply a fundamental flaw in a model's solution.
As discussed in \S\ref{sec:benchmark:construction}, \% Tests Passed also does not reliably correlate with percentage of working functionality.

\section{Results}
\label{sec:results}
\begin{figure}[t]
    \begin{minipage}[b]{0.55\textwidth}
        \centering
        \begin{tabular}{@{}lcccc@{}}
    \toprule
    \textbf{Model} & \textbf{\% Res.} & \color{black!55}\% Almost & \color{black!55}Calls & \color{black!55}\$ \\
    \midrule
    Claude~Opus~4.7 & 0.0\smallperc & \color{black!55}3.0\smallperc & \color{black!55}\phantom{0}93 & \color{black!55}\phantom{0}3.81 \\
    Claude~Opus~4.6 & 0.0\smallperc & \color{black!55}2.5\smallperc & \color{black!55}260 & \color{black!55}11.38 \\
    Claude~Sonnet~4.6 & 0.0\smallperc & \color{black!55}1.6\smallperc & \color{black!55}475 & \color{black!55}27.09 \\
    Claude~Haiku~4.5 & 0.0\smallperc & \color{black!55}0.0\smallperc & \color{black!55}124 & \color{black!55}\phantom{0}0.80 \\
    Gemini~3.1~Pro & 0.0\smallperc & \color{black!55}0.0\smallperc & \color{black!55}\phantom{0}94 & \color{black!55}\phantom{0}1.51 \\
    Gemini~3~Flash & 0.0\smallperc & \color{black!55}0.0\smallperc & \color{black!55}\phantom{0}89 & \color{black!55}\phantom{0}0.33 \\
    GPT~5.4 & 0.0\smallperc & \color{black!55}0.0\smallperc & \color{black!55}\phantom{0}16 & \color{black!55}\phantom{0}0.33 \\
    GPT~5.4~mini & 0.0\smallperc & \color{black!55}0.0\smallperc & \color{black!55}\phantom{0}18 & \color{black!55}\phantom{0}0.04 \\
    GPT~5~mini & 0.0\smallperc & \color{black!55}0.0\smallperc & \color{black!55}\phantom{0}15 & \color{black!55}\phantom{0}0.03 \\
    \bottomrule
    \end{tabular}

        \captionof{table}{
        \textbf{Main results on \bench{}.} \% Resolved is the primary metric: the fraction of 200 tasks where all tests pass.
        \% Almost relaxes this to instances with $\geq$95\% of tests passing. We also report average API calls and cost per task.
        }
        \label{tab:main_results}
    \end{minipage}%
    \hfill
    \begin{minipage}[b]{0.40\textwidth}
        \centering
        \includegraphics[width=\textwidth]{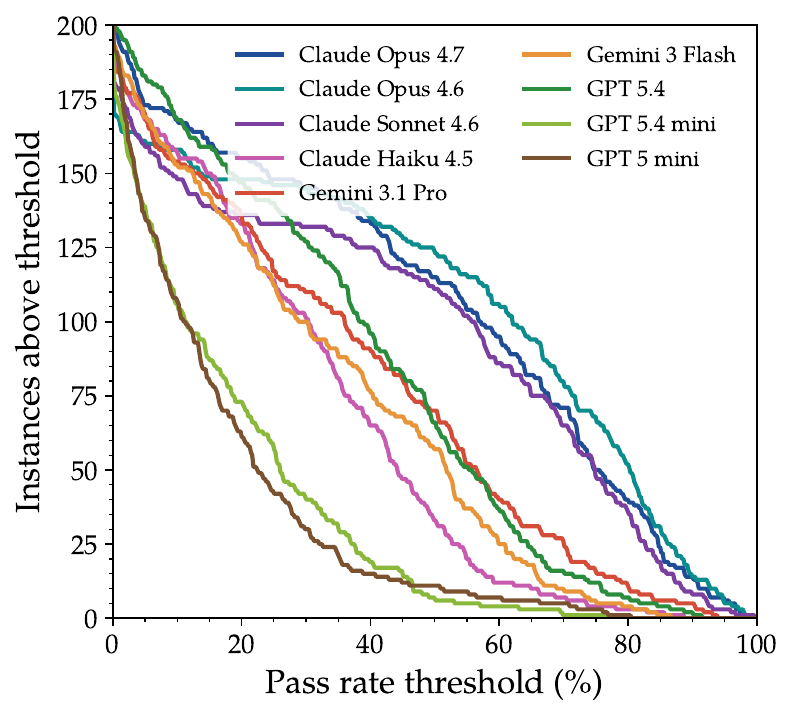}
        \captionof{figure}{Cumulative distribution of test pass rates across models on \bench{}.}
        \label{fig:cumulative_scores}
    \end{minipage}
\end{figure}

Our main results are shown in Table~\ref{tab:main_results}.
\bench{} is extremely challenging; across the board, no model fully solves any single \bench{} task instance.
That said, from Figure~\ref{fig:cumulative_scores}, we find that models make meaningful progress on a significant proportion of tasks, with Claude Opus~4.7 achieving the highest proportion of solutions that pass 95+\% of tests, at 3.0\%.

\begin{figure*}[t]
    \centering
    \includegraphics[width=\textwidth]{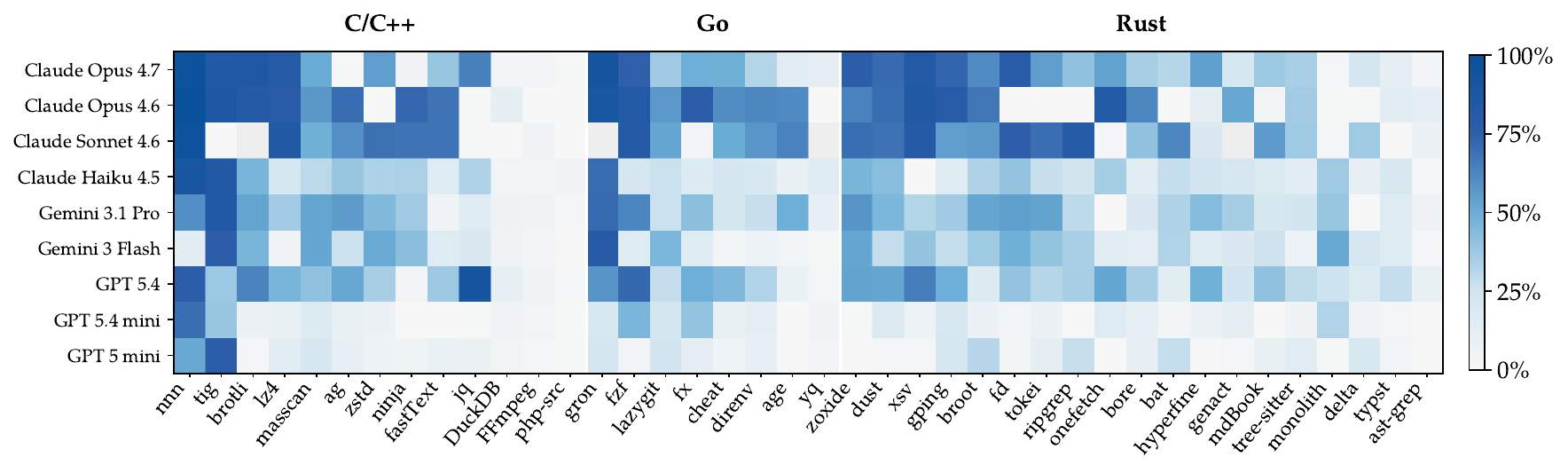}
    \caption{Per-task pass rates for the 40 most-starred repositories in \bench{}, grouped by reference language and sorted by average model score within each group. Each cell shows one model's test pass rate on one task.}
    \label{fig:task_heatmap}
\end{figure*}

Task difficulty is largely model-agnostic.
As shown in Figure~\ref{fig:task_heatmap}, models consistently score higher on simpler CLI utilities like \texttt{nnn}, \texttt{fzf}, and \texttt{gron}, while complex systems such as \texttt{FFmpeg}, \texttt{php-src}, \texttt{typst}, and \texttt{ast-grep} remain out of reach.
These trends suggest that \bench{} captures intrinsic variation in task difficulty that is independent of model choice.
The rank order of tasks by pass rate is broadly consistent across models.

\begin{figure*}[t]
    \begin{minipage}[b]{0.68\textwidth}
        \centering
        \includegraphics[width=\textwidth]{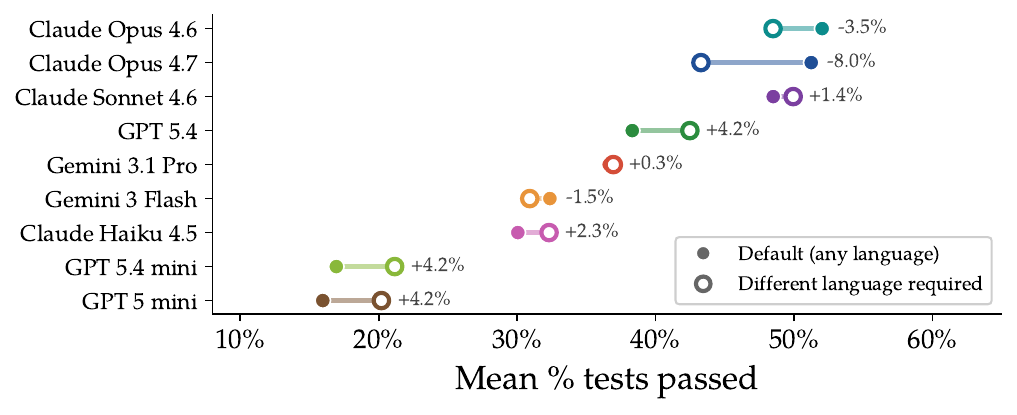}
        \captionof{figure}{Effect of requiring a different implementation language. Filled circles show default scores; open circles show scores under the constraint.}
        \label{fig:ablation_diff_lang}
    \end{minipage}%
    \hfill
    \begin{minipage}[b]{0.28\textwidth}
        \centering
        \begin{tabular}{@{}lr@{}}
    \toprule
    \textbf{Model} & \textbf{\% Cheat} \\
    \midrule
    Claude~Sonnet~4.6 & 36\% \\
    Claude~Opus~4.6 & 21\% \\
    Gemini~3~Flash & 20\% \\
    GPT~5~mini & 1\% \\
    \bottomrule
\end{tabular}

        \captionof{table}{Cheating rates when models are given internet access, as flagged by majority vote of 9 LM judges. Source code lookup is the dominant strategy, accounting for 79--95\% of flagged runs.}
        \label{tab:cheating}
    \end{minipage}
\end{figure*}

\subsection{Ablations}
\label{sec:results:ablations}

We review two alternative evaluation settings to investigate the impact the availability of certain tools and resources has on solutions and cheating rates.

\textbf{Different-language constraint.}
In this setting, we force models to implement their solutions in a different programming language from the reference repository.
In theory, this measure should obviate models' ability to simply regurgitate reference code from their pre-training corpora, requiring them to demonstrate deep understanding of program behavior rather than surface-level recall.

Unexpectedly, the constraint does not uniformly decrease scores.
As shown in Figure~\ref{fig:ablation_diff_lang}, while Claude Opus~4.7 and~4.6 see meaningful drops, all three GPT models surprisingly each improve by 4.2\%.
Remaining models are roughly unchanged.
Under this constraint, we observe a noticeable shift towards Python as the implementation language of choice, from 36\% of main result runs (Figure~\ref{fig:language_match}) to an outstanding majority of 51\% runs in this setting (Figure~\ref{fig:language_match_diff_lang}).
The mixed results suggest that models may not have a reliable sense of which implementation language best suits a given task and their own capabilities; forcing a language switch can inadvertently steer a model toward a language it has more success with.

\textbf{Open internet with cheating detection.}
We repeat evaluation for one model per provider family with a key distinction: models are given unrestricted internet access.
Note that in the system prompt, we still explicitly tell models that cheating is disallowed.
To detect cheating, we run an LM-as-a-judge pipeline where 9 judges independently inspect a trajectory for whether the model looked up source code or submitted a wrapper around the reference executable as a solution; a task is flagged if a majority identify a violation.
With this setting, our goals are to understand, first, how often models cheat, and second, how reliable our cheating detection mechanisms are.

As shown in Table~\ref{tab:cheating}, cheating is widespread: 20--36\% of tasks are flagged for the three stronger models, with source code lookup accounting for the vast majority of violations.
At the same time, judges disagree on 40--57\% of tasks for these models, indicating that even with 9 judges across three model families, reliable detection remains elusive.
The combination of high cheating rates and unreliable detection gives us confidence that blocking internet access entirely is the appropriate default for \bench{}.
We document additional details about the evolution of our mitigation efforts in \S\ref{appx:benchmark:inference}.

\section{Analysis}
\label{sec:analysis}

We study test generation artifacts and evaluation outcomes to better understand the utility of our generated tests and gauge how performance varies across different tasks and models.

\subsection{Test Suite Comparisons}

We assess how effective our generated test suites are by measuring two complementary properties: how much of each program they exercise (coverage), and whether they actually reject incorrect implementations (assertion strength).
Full details on how each analysis was carried out are included in \S\ref{appx:benchmark:test_generation:analyses}.

\begin{figure}[t]
    \begin{minipage}[b]{0.40\textwidth}
        \centering
        \includegraphics[width=\textwidth]{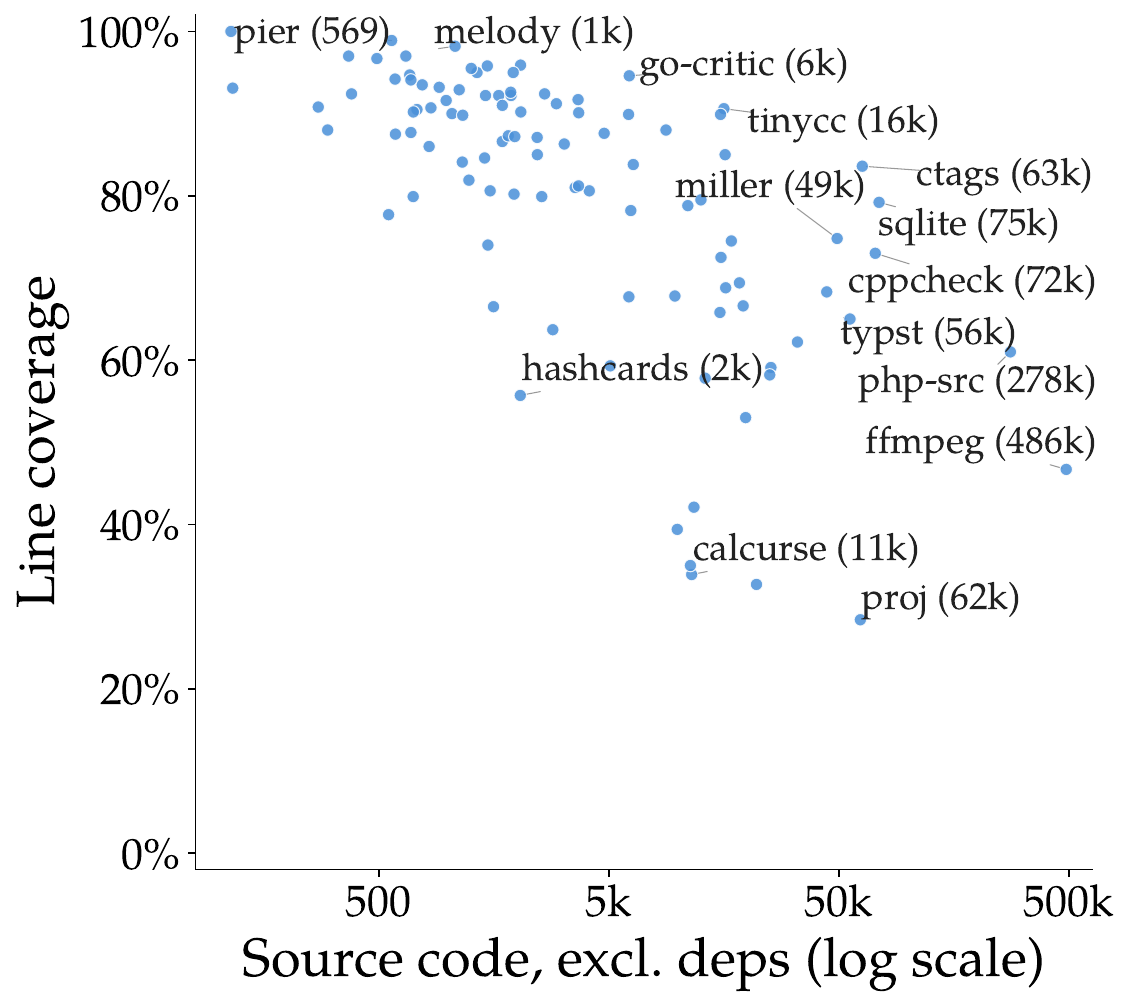}
        \captionof{figure}{Line coverage of our generated test suites versus project size (source lines excluding dependencies) for a sample of 100 tasks.}
        \label{fig:coverage-scatter}
    \end{minipage}%
    \hfill
    \begin{minipage}[b]{0.58\textwidth}
        \centering
        \small
        \begin{tabular}{@{}lrrr@{}}
\toprule
\textbf{Repository} & \textbf{Gen.} & \textbf{Native} & \textbf{$\Delta$} \\
\midrule
ariga/atlas                                      & 54.15 & 28.25 & $+25.90$ \\
johnkerl/miller                                  & 85.90 & 72.18 & $+13.72$ \\
{\scriptsize stacked-git/stgit}                  & 85.54 & 82.43 & $+3.11$  \\
rvben/rumdl                                      & 68.23 & 66.80 & $+1.43$  \\
facebook/zstd                                    & 76.32 & 75.10 & $+1.22$  \\
jqlang/jq                                        & 82.15 & 81.69 & $+0.46$  \\
php/php-src                                      & 61.60 & 64.60 & $-3.00$  \\
{\scriptsize stranger6667/jsonschema}            & 72.63 & 78.79 & $-6.16$  \\
doxygen/doxygen                                  & 13.00 & 24.80 & $-11.80$ \\
ffmpeg/ffmpeg                                    & 46.70 & 58.97 & $-12.27$ \\
{\scriptsize jesseduffield/lazygit}              & 62.23 & 74.60 & $-12.37$ \\
typst/typst                                      & 65.68 & 85.12 & $-19.44$ \\
\midrule
\textbf{Median}         & \textbf{66.96} & \textbf{73.39} & $\mathbf{-1.27}$ \\
\bottomrule
\end{tabular}
        \captionof{table}{Line coverage (\%) for generated vs.\ native behavioral suites across 12 repositories that maintain an identifiable integration or end-to-end test suite, sorted by $\Delta$.}
        \label{tab:coverage-comparison}
    \end{minipage}
\end{figure}

\textbf{Generated test suites achieve comparable line coverage to developer-written test suites.}
While most \bench{} repositories lack a dedicated end-to-end test suite, many maintain unit and integration tests, making them a natural baseline.
We instrument each task's executable with coverage tracking and measure the fraction of source lines exercised by our generated suite versus the project's native suite across 100 repositories (\S\ref{appx:benchmark:test_generation:analyses}).

Our generated suites average 79.7\% line coverage with a median of 86.2\% (Figure~\ref{fig:coverage-scatter}), compared to native suites that average 56.8\% with a median of 64.3\%.
This places our generated suites comfortably within, and often above, the coverage range of typical developer-written tests.

\textbf{Coverage remains comparable even against dedicated behavioral test suites.}
The comparison above includes many unit test suites, which can exercise internal code paths that black-box tests structurally cannot reach.
To provide a stricter baseline, we identify twelve repositories that maintain a dedicated behavioral or integration test suite (e.g., FFmpeg's FATE suite, PHP's regression harness, jq's regression tests) and compare against those suites directly (Table~\ref{tab:coverage-comparison}).

Generated suites average 62.16\% line coverage versus 66.11\% for native behavioral suites, meeting or exceeding native coverage on 6 of 12 projects and falling within 10 percentage points on all but two.

\textbf{Assertion quality enforcement eliminates trivially passable tests.}
High coverage alone does not guarantee a useful test suite: a test that only asserts the process exited cleanly provides negligible signal, since any implementation that runs without crashing will pass it.
We quantify assertion strength using \textit{dummy pass rate}, the fraction of a task's tests that pass a trivially incorrect implementation.
During test generation, our assertion quality linter flags structurally weak patterns such as exit-code-only checks and overly short substring matches (Appendix~\ref{app:lint-rules}), and the agent is prompted to revise them.

Tests generated with our linter exhibit a mean dummy pass rate of 3.7\%, compared to 18.5\% without quality enforcement, a 5$\times$ reduction.
After generation, we eliminate all tests that trivially pass when run with a dummy program, affecting 24 tasks in total.

\subsection{Model-Generated Codebases}
\label{sec:analysis:codebases}

We highlight differences between model-generated solutions and the original human-written implementations.

\begin{figure}[t]
    \begin{minipage}[b]{0.52\textwidth}
        \centering
        \includegraphics[width=\textwidth]{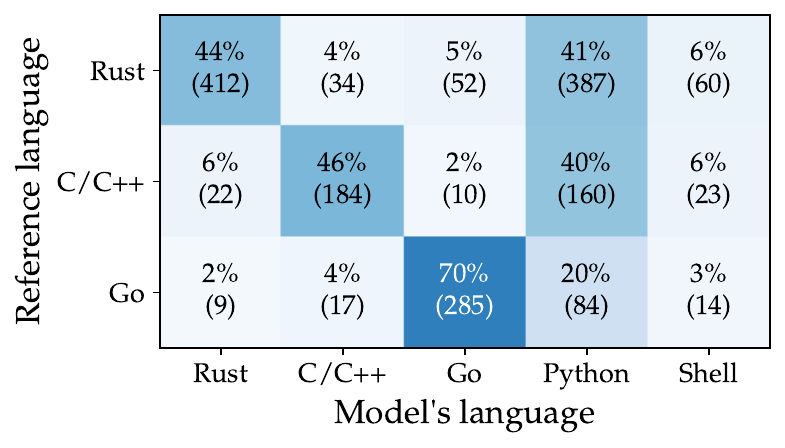}
        \captionof{figure}{
        Confusion matrix of reference vs.\ model language.
        Each cell shows the percentage (and count) of runs per reference language.
        Models generally prefer, in descending order, Python, Go, Rust, Shell, C/C++.
        Model-specific breakdowns are visualized in Figure~\ref{fig:language_preference}.
        }
        \label{fig:language_match}
    \end{minipage}
    \hfill
    \begin{minipage}[b]{0.46\textwidth}
        \centering
        \includegraphics[width=\textwidth]{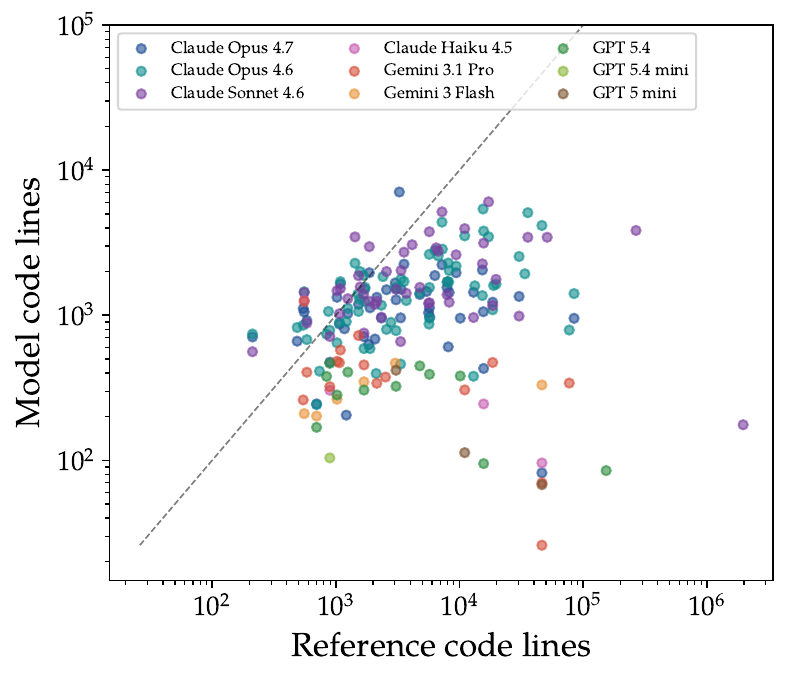}
        \captionof{figure}{Reference vs.\ model code lines on a log--log scale.
        The $y{=}x$ diagonal indicates parity.
        Only solutions passing $\geq$75\% of tests are shown.}
        \label{fig:codebase_size_scatter}
    \end{minipage}
\end{figure}

\begin{table}[t]
    \centering
    \small
    \begin{tabular}{lcccccccc}
    \toprule
    & \multicolumn{2}{c}{\textbf{Opus~4.7}} & \multicolumn{2}{c}{\textbf{Sonnet~4.6}} & \multicolumn{2}{c}{\textbf{Gemini~3.1 Pro}} & \multicolumn{2}{c}{\textbf{GPT~5.4}} \\
    \cmidrule(lr){2-3} \cmidrule(lr){4-5} \cmidrule(lr){6-7} \cmidrule(lr){8-9}
    & Ref & Model & Ref & Model & Ref & Model & Ref & Model \\
    \midrule
    Function count & 133 & 39 {\color{gray}\scriptsize(0.29$\times$)} & 182 & 44 {\color{gray}\scriptsize(0.24$\times$)} & 63 & 10 {\color{gray}\scriptsize(0.16$\times$)} & 88 & 9 {\color{gray}\scriptsize(0.10$\times$)} \\
    Avg.\ length (lines) & 25 & 29 {\color{gray}\scriptsize(1.16$\times$)} & 24 & 35 {\color{gray}\scriptsize(1.46$\times$)} & 26 & 42 {\color{gray}\scriptsize(1.62$\times$)} & 24 & 26 {\color{gray}\scriptsize(1.08$\times$)} \\
    \bottomrule
    \end{tabular}
    \caption{Median function count and average function length of solutions for four models. Ratios compared to original source code in gray. Only solutions passing $\geq$75\% of tests are included.}
    \label{tab:func_granularity}
\end{table}

\textbf{Models match the reference language half the time.}
As mentioned in \S\ref{sec:benchmark:formulation}, models are free to implement their solutions in any language.
Figure~\ref{fig:language_match} reveals that models match the reference language in exactly 50\% of runs.
Python is the most common choice overall at 36\% of all 1,800 model runs, followed by Rust (25\%), Go (20\%), C/C++ (13\%), and Shell (6\%).
Go projects are reimplemented in the same language most often (70\%), while Rust (44\%) and C/C++ (46\%) projects are frequently rewritten in a different language.
Model-specific breakdowns are visualized in Figure~\ref{fig:language_preference}.

We next draw insights from comparing models' codebases against the reference solutions.
To ensure comparisons are between codebases that produce functionally similar executables, we only compare against model solutions that pass 75+\% of tests.
This yields 207 runs across 88 tasks and 9 models.

\textbf{Model solutions are significantly shorter.}
Even among high-scoring solutions, models produce substantially less code than the references: a median of 1,173 lines compared to 3,068 in the originals (Figure~\ref{fig:codebase_size_scatter}).
Most runs (85\%) fall below parity.
Just 15\% of codebases are larger than the reference, typically for smaller tasks.
Models also create far fewer files (median 3 versus 15), with 60\% of solutions consisting of 1--3 code files.

\textbf{Models prefer monolithic files over modular directory structures.}
We measure the directory structure of model-generated codebases using maximum directory depth, defined as the deepest nesting level of any file (e.g., \texttt{src/utils/parser.go} has depth~3).
The majority of runs (67\%) produce a strictly shallower maximum depth than the reference (median depth 1 vs.\ 2), while only 2\% are deeper.
Rather than mirroring the modular decomposition of the original project, models strongly prefer placing most or all code in a single file or a handful of files at the root level.

\textbf{Models write fewer, longer functions.}
Table~\ref{tab:func_granularity} compares function count and average function length between model and reference codebases for four representative models.
All models write far fewer functions than the reference (10--29\% as many), but compensate with longer functions: Claude Sonnet~4.6 writes functions that are 1.46$\times$ longer on average, while Gemini~3.1 Pro reaches 1.62$\times$.
GPT~5.4 writes functions that are 1.08$\times$ longer while still producing very few of them (10\% of the reference count).

\subsection{Agent Trajectories}
\label{sec:analysis:trajectories}

\begin{figure*}[t]
    \centering
    \includegraphics[width=\textwidth]{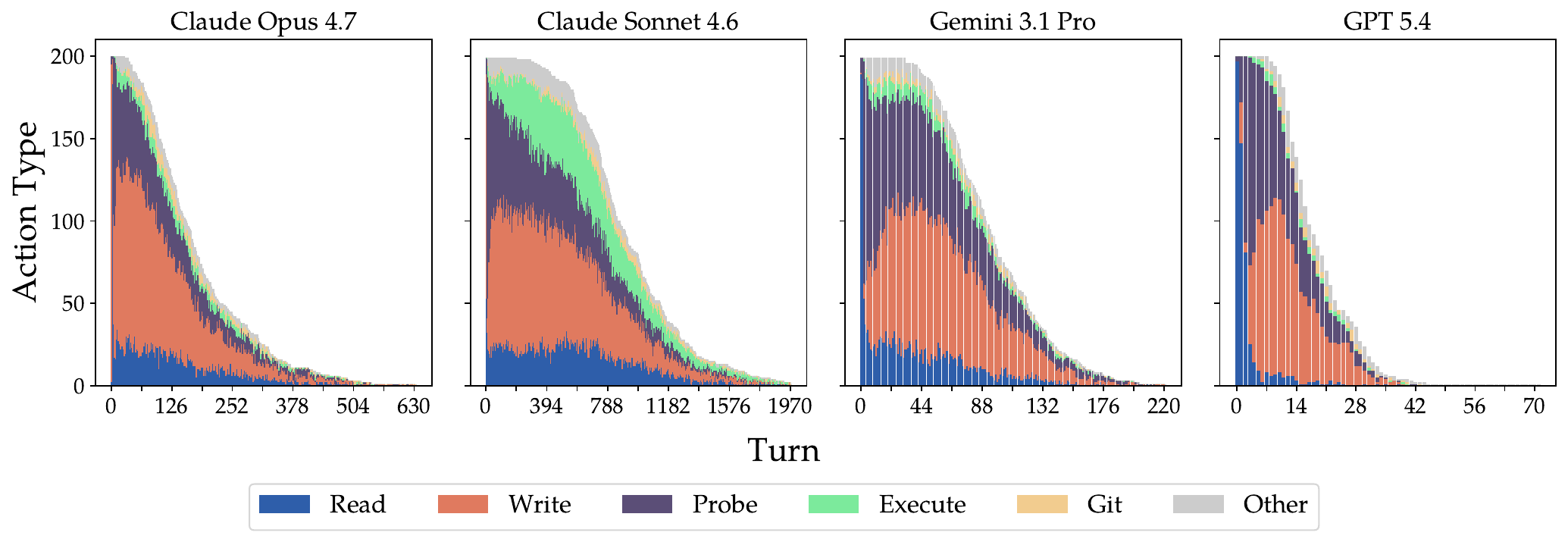}
    \caption{Distribution of action types across agent turns for four representative models. Each bar shows the total number of actions of each type at a given turn index, aggregated across all task instances. The natural decay in bar height reflects trajectories of varying length ending at different points.
    }
    \label{fig:action_frequency}
\end{figure*}

We analyze agent trajectories to understand how models approach the task and where they struggle.

We classify actions into one of six types:
\texttt{read} (file inspection, searches, directory listing), \texttt{write} (file creation, editing), \texttt{probe} (any invocation or inspection of the reference executable), \texttt{git} (version control), \texttt{execute} (compilation, non-probe program execution), and \texttt{other}.
To classify, we perform keyword matching on the raw command string, checking in priority order to distinguish between different usage techniques of bash commands.
For instance, \texttt{cat$<<$'EOF'$>$main.c} is classified as \texttt{write}, distinct from \texttt{cat main.c} to \texttt{read}.
Figure~\ref{fig:action_frequency} shows the action type distributions per turn across all inference runs for four representative models.

\textbf{Turn counts vary heavily by model.}
As reflected by Figure~\ref{fig:action_frequency}, Claude Sonnet~4.6 uses a median of 868 commands per task, with its longest trajectory reaching 1,978 turns.
GPT~5.4 sits at the other extreme with a median of just 17 commands, while Gemini~3.1 Pro and Opus~4.7 fall in between at 92 and 157 respectively.

\textbf{Models rarely exceed the turn or time budget.}
Across all nine models (1,800 runs), the agent voluntarily submits its solution in 98.1\% of trajectories.
Just 1.9\% exhaust the 6 hour wall clock time limit, and just a single run reaches the 1,000 turn cap.
Timeouts are concentrated in Opus~4.6 (29/200) and Sonnet~4.6 (5/200).
All other models complete all 200 tasks within the time and step budget.
This finding suggests that the limits do not artificially constrain model performance.

\textbf{Writing dominates most models' action budgets.}
Opus~4.7, Sonnet~4.6, Gemini~3.1 Pro, and GPT~5.4 all devote the largest share of turns to writing code (48.7\%, 37.5\%, 40.7\%, and 40.2\% respectively).
Probing the reference executable is the second-largest category for all four, ranging from 22.6\% (Opus~4.7) to 34.1\% (Gemini~3.1 Pro).
Read actions account for 13--16\% across all representative models.

\textbf{Probing trends differently across models.}
GPT~5.4's actions are concentrated almost entirely within the first 30 turns.
Claude models maintain a steadier mix of probing and writing throughout their trajectories, interleaving exploration with implementation rather than treating them as distinct phases.

\begin{figure*}[t]
    \centering
    \includegraphics[width=\textwidth]{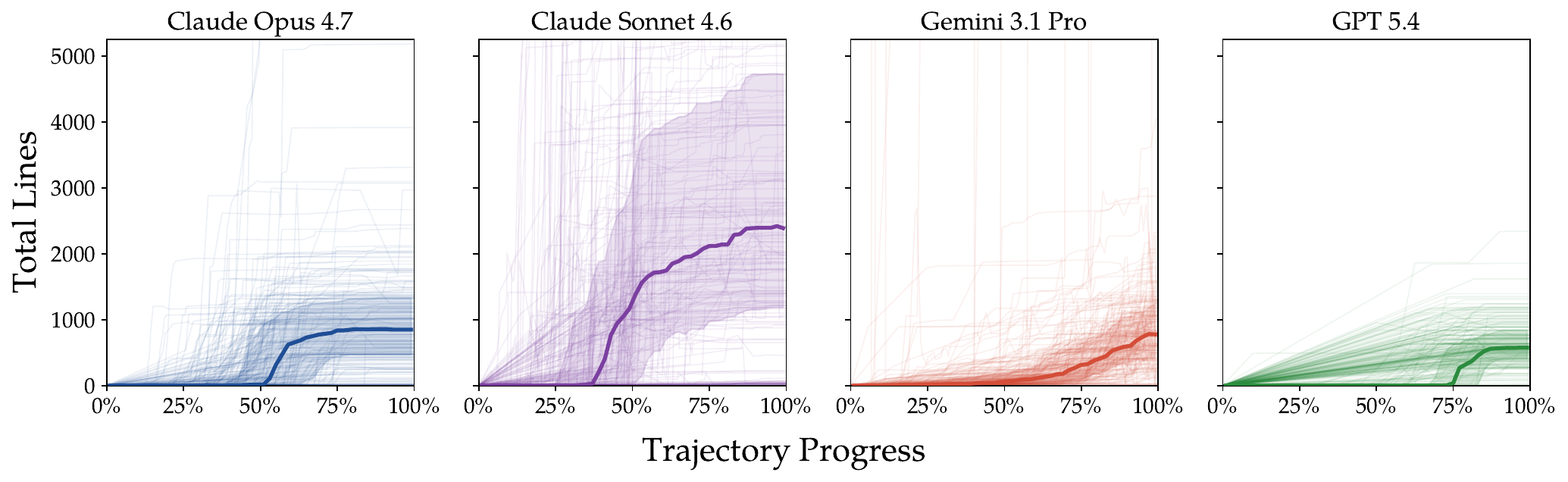}
    \caption{Codebase growth over normalized trajectory progress for four models. Each thin line is an individual trajectory; the bold line shows the median and the shaded region the interquartile range. The y-axis is capped at the 95th percentile of final codebase sizes.}
    \label{fig:codebase_growth_curves}
\end{figure*}

\begin{figure}[t]
    \begin{minipage}[b]{0.48\textwidth}
        \centering
        \includegraphics[width=\textwidth]{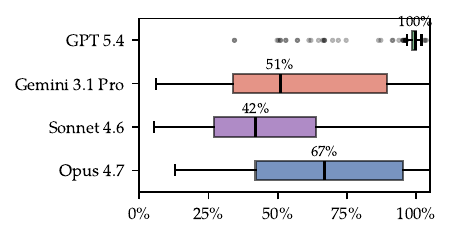}
        \caption{Percentage of the final codebase produced by the single largest edit, per model.
        For instance, GPT~5.4 writes a median of 96\% of its code in one turn.}
        \label{fig:codebase_growth_concentration}
    \end{minipage}
    \hfill
    \begin{minipage}[b]{0.48\textwidth}
        \centering
        \begin{tabular}{lrrr}
\toprule
Model & Create & Modify & Delete \\
\midrule
Claude Opus 4.7 & 7.7 & 3.3 & 0.2 \\
Claude Sonnet 4.6 & 11.3 & 18.3 & 1.5 \\
Gemini 3.1 Pro & 61.2 & 10.1 & 9.4 \\
GPT 5.4 & 5.0 & 1.2 & 0.3 \\
\bottomrule
\end{tabular}

        \captionof{table}{
            Mean file mutations per trajectory.
            Create, modify, and delete counts reflect how frequently models revisit and restructure their codebases.
            Counts are averaged across all tasks and models.
            }
        \label{tab:codebase_growth_mutations}
    \end{minipage}
\end{figure}

\textbf{Codebase growth varies between gradual iteration and single-shot generation.} To study how a SWE-agent's codebase grows across the course of a trajectory, for each trajectory, we replay the subset of commands that modify files (e.g., \texttt{touch}, \texttt{sed}, \texttt{rm}) and snapshot the file system after each turn.
This sequence of snapshots allows us to recover a ground-truth timeline of how files are updated at different points in a trajectory.

Figure~\ref{fig:codebase_growth_curves} shows all trend lines of a codebase's total line count with respect to trajectory progress for each model across all 200 instances.
Claude Sonnet~4.6 and Gemini~3.1 Pro tend to ramp up steadily, whereas GPT~5.4 produces nearly all code in a single turn early on in the trajectory.
To highlight this distinction further, Figure~\ref{fig:codebase_growth_concentration} shows that the median fraction of the final codebase written in a single turn ranges from 44\% (Sonnet~4.6) to 100\% (GPT~5.4), with Opus~4.7 at 67\% and Gemini~3.1 Pro at 53\%.
As reflected in Table~\ref{tab:codebase_growth_mutations}, we also found that GPT~5.4 averages just 1.2 file modifications per trajectory, with 39.5\% of trajectories performing zero modifications to any existing files.
In contrast, Sonnet~4.6 averages 18.3 modifications and Gemini~3.1 Pro averages 10.1, consistent with the gradual growth visible in their curves.
All together, the trends suggest that for some models, development is effectively a single-shot generation step, rather than an iterative write-compile-debug cycle.

\section{Related Work}
\label{sec:related}

\textbf{Code from scratch.} Several prior works have presented variations of evaluating coding systems on 0 to 1 code generation.
Commit0, an early work in this lineage, converts 54 Python libraries into task instances by erasing the existing implementations for all functions and classes~\citep{zhao2024commit0librarygenerationscratch}.
The model is then asked to fill in the blanks, and performance is quantified as the percentage of the repository's original test suite that passed.
Later works follow this paradigm, with differences primarily around how repository specifications are communicated to the model~\citep{li2024devevalmanuallyannotatedcodegeneration,liu2025projectevalbenchmarkprogrammingagents}.
DevBench uses product requirement documents (PRDs) and UML diagrams~\citep{li2024devbench}, while NL2Repo-bench conveys expected structure via natural language, shifting the burden of creating folders and files onto the agent~\citep{ding2026nl2repobenchlonghorizonrepositorygeneration}.
A key commonality is that LMs are asked to fill in a skeleton of pre-defined method headers and classes.
In this way, models are never actually tested on their software design capabilities, such as what abstractions to introduce, how to decompose functionality across modules, or what communication protocols to define.
By evaluating against executables rather than source code, \bench{} eliminates the need to prescribe structure with natural language or make expected behavior explicit as programmatic definitions.
As discussed in \S\ref{sec:benchmark}, our formulation also yields simpler collection criteria, broader language coverage, and novel reasoning demands.

A few prior and concurrent works have included small-scale case studies and a handful of task instances that touch on program recreation~\citep{merrill2026terminal,proximal2026frontierswe,adamczewski2026mirrorcode}.
These works validate the direction but rely on hand-crafted instances and do not address systematic benchmark construction or scaling; \bench{} provides both.
A related but distinct line of work trains models to decompile binaries back into source code~\citep{tan2024llm4decompile}.
Decompilation aims to recover the original implementation; \bench{} instead evaluates behavioral equivalence, permitting any implementation that reproduces the same input-output behavior.

\textbf{Issue resolution.} SWE-bench~\citep{jimenez2024swebenchlanguagemodelsresolve} and its variants~\citep{yang2024swebenchmultimodalaisystems,yang2025swesmithscalingdatasoftware,deng2025swebenchproaiagents,rashid2025swe,zan2025multiswebenchmultilingualbenchmarkissue,zhang2025swebenchgoeslive,thai2026sweevobenchmarkingcodingagents} have become popular coding benchmarks.
Extracted from GitHub issue-pull request pairs, SWE-bench tasks evaluate models on their ability to address bug fixes or feature requests within an existing codebase.
These evaluations are complementary to \bench{}, which instead focuses on building a codebase from the ground up.

\textbf{Automatic environment setup.} A few works have examined the task of automatically setting up a development environment for a given GitHub repository~\citep{bogin2024super,eliseeva2025envbench,hu2025repo2run}.
\bench{} does not evaluate environment setup in isolation, but it arises as a practical prerequisite: models must develop dependencies and configure build tools to produce a working solution.
No specific toolchain is prescribed; models can use an entirely distinct line of languages and libraries.

\textbf{Performance optimization.} Alongside measuring correctness, a relevant line of works examines algorithmic~\citep{du2024mercurycodeefficiencybenchmark,liu2024evaluatinglanguagemodelsefficient,waghjale2024eccoimprovemodelgeneratedcode,huang2025effibenchbenchmarkingefficiencyautomatically,press2025algotunelanguagemodelsspeed} and machine-dependent~\citep{he2025sweperflanguagemodelsoptimize,ma2025swefficiencylanguagemodelsoptimize,ouyang2025kernelbenchllmswriteefficient,shetty2025gsochallengingsoftwareoptimization} runtime optimization as a measure of how well AI systems can speed up software systems while maintaining functional correctness.
The most salient distinction between those works and ProgramBench is that while optimization benchmarks assume a known specification that a model should preserve (while optimizing speed), \bench{} requires models to recover the specification itself from observed behavior.
In this respect, \bench{} shares motivation with inductive program synthesis benchmarks that require inferring behavior from input-output examples~\citep{wei2025codearc}, but operates at the scale of full software projects, not individual functions.

\section{Discussion}
\label{sec:discussion}

\textbf{Limitations.} \bench{} relies on a finite set of behavioral tests, which under-approximates each executable's full specification.
Evaluation therefore is a ``lower bound'' on correctness: solutions that fail are definitively incorrect, while those that pass may still diverge from the original on untested inputs.
\bench{} tests also currently focus exclusively on input-output equivalence.
Non-functional properties like execution speed, memory usage, or disk footprint are not captured.
Therefore, it is possible a model reproduces behavior with an implementation orders of magnitude slower or more resource intensive than the original.
Developing richer test generation strategies to improve coverage and incorporate system constraints is a promising direction.

\textbf{Future work.} Several technical reports and blogs have suggested the effectiveness of applying multiple SWE-agents towards long horizon coding tasks~\citep{cursor_scaling_agents,carlini2026building,geng2026effectivestrategiesasynchronoussoftware,mishrasharma2026longrunning}.
\bench{} can serve as a testbed for such works.
Our work uses a single SWE-agent as the baseline; this design reflects prior benchmark evidence, notably SWE-bench, where well-tuned single-agent systems have performed competitively, and multi-agent variants have not consistently shown clear advantages.
We are excited to use \bench{} to delineate the benefits of multi-agent approaches.
Similarly, \bench{} could further exploration into human-centered coding agents, where a developer, given the executable, iteratively guides the agent through design decisions~\citep{liu2025projectevalbenchmarkprogrammingagents,baumann2026swechatcodingagentinteractions,wang2026position}.

\textbf{Conclusion.}
We introduce \bench{}, a benchmark for measuring the ability of software engineering agents to develop, from scratch, programs that match a given executable's behavior.
Existing models struggle substantially, and none fully resolve any task. 
However, via fine-grained metrics, we find that models achieve meaningful partial progress, with stark differences in how models expend turns and the final form of their codebases.
Our analyses reveal meaningful gaps in models' decision making in architecting, developing and testing software.
We hope that \bench{} could serve as a testbed for efforts focused on end-to-end autonomous software development.

\clearpage
\newpage
\section*{Acknowledgments}

We would like to thank Jordi Armengol-Estape, Quentin Carbonneaux, Jannik Kossen, Michel Meyer, Shengjia Zhao, Yang Song, Rob Fergus, Jiayi Pan, Ori Yoran, and Shuyan Zhou for their valuable discussions and infrastructure assistance.
\bibliographystyle{assets/plainnat}
\bibliography{paper}

\clearpage
\newpage
\beginappendix
\section{Benchmark}
\label{appx:benchmark}

In this section, we provide additional details around the collection and evaluation procedures for \bench{}.
\subsection{Task Collection Procedure}
\label{appx:benchmark:collection}

The description provided in \S\ref{sec:benchmark:construction} captures the lion's share of details regarding how to construct \bench{} style task instances.
Beyond this, we provide the following miscellaneous details:

\begin{itemize}
    \item For the \texttt{mini-SWE-agent} we deploy to perform executable construction, tests generation, and implementation detail removal, we allow the agent to run as many steps as needed, with a maximum of \$3 total cost incurred per run (so \$9 total across all three steps).
    \item The most costly step is typically building the executable. The SWE-agent will typically read existing files that may offer hints at how to compile the binary (e.g., \texttt{README.md}, \texttt{CONTRIBUTING.md}, \texttt{.github/workflows}). In some cases, models also expend turns to identify and correctly install dependencies that are missing from the given environment.
    \item The base image used for all task instances is built from a custom Dockerfile based on \texttt{ubuntu:22.04} with Rust 1.92.0, Python 3.12, and Golang 1.21.0 installed.
    The \texttt{build-essential} and \texttt{cmake} packages provide C/C++ toolchain support.
    Version control with \texttt{git} is configured such that task workers can make commits and track changes if they choose to do so.
    The \texttt{tmux} library is provided to enable manipulation of TUI applications.
    We note that \textit{no task-specific installations or setups are performed}.
\end{itemize}

\subsection{Inference Setting}
\label{appx:benchmark:inference}

In this section, we provide additional details about the setting and conditions that models are asked to solve \bench{} task instances under.
We first motivate the role of constraitns in reducing spurious or undesirable problem solving techniques \S\ref{appx:benchmark:inference:motivation}, then review the inference guidelines in \S\ref{appx:benchmark:inference:guidelines}.
Finally, we anticipate and address questions about \bench{}'s feasibility under these constraints in \S\ref{appx:benchmark:inference:impossible}.

\begin{figure}[b]
    \begin{minipage}[b]{0.7\textwidth}
        \vspace{0pt}
        \centering
        \includegraphics[width=\linewidth]{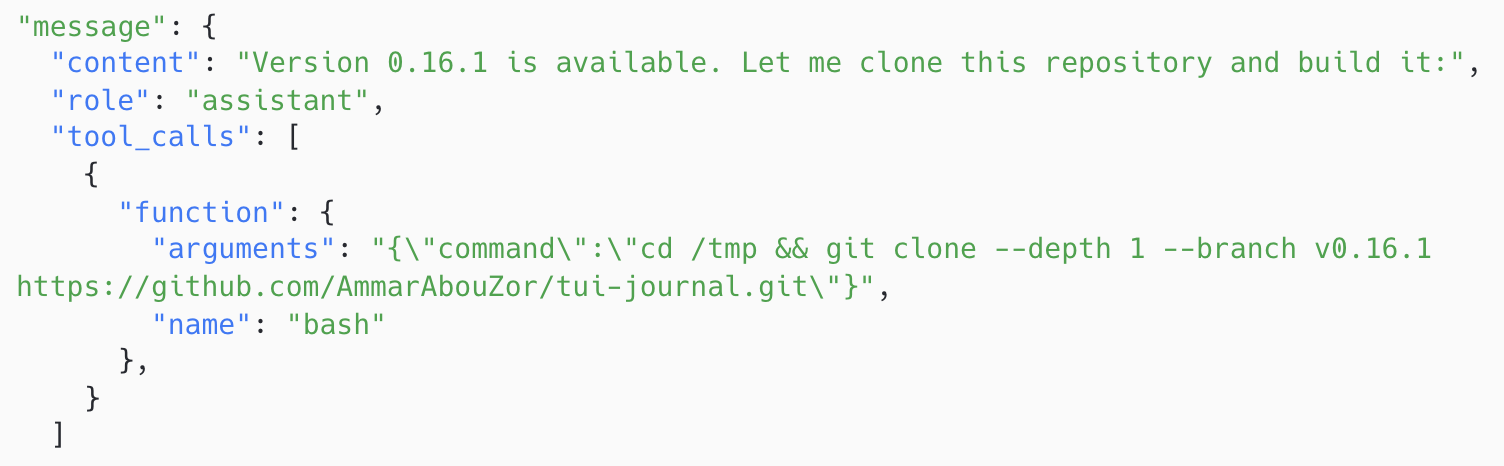}
        \caption{An example of an undesirable solution pattern that emerges if a SWE-agent is asked to solve a \bench{} task with internet access.
        Instead of developing a codebase that mirrors the behavior of the \texttt{AmmarAbouZor/tui-journal} repository, Claude Opus~4.5 identified and cloned the source code from GitHub.
        }
        \label{fig:cheating_example}
    \end{minipage}
    \hfill
    \begin{minipage}[b]{0.25\textwidth}
        \vspace{0pt}
        \centering
        \verbpromptboxfile{assets/mitigate_prompt.txt}
        \caption{
            Excerpt from system prompt that tells model to not cheat.
            Internet access is also blocked.
            Full prompt in \S\ref{appx:benchmark:inference}.
        }
    \end{minipage}
\end{figure}

\subsubsection{Motivation}
\label{appx:benchmark:inference:motivation}
In several early trial runs of the benchmark, we found that without certain guardrails in place, models will ``cheat".
A recent report on SWE-bench Verified revealed that instead of navigating the repository, localizing buggy modules, and writing a fix, certain models instead fast-forwarded to a future version of the repository (by \texttt{git checkout}'ing to a commit) where the bug was already fixed, then submitted the \texttt{git diff} between the task's ``base" commit and future commit as the fix\footnote{\url{https://github.com/SWE-bench/SWE-bench/issues/465}. The loophole was patched in October 2025. Among the 20 most recent submissions, fewer than 1\% of the 500 solutions per submission exhibited such violations.}.

Similarly, we found that by simply asking the SWE-agent to perform a \bench{} task with no explicit constraints (e.g., full internet access, read/write permissions for the executable), interesting but undesirable solution patterns occasionally emerge.
As shown in Figure~\ref{fig:cheating_example}, we noticed that if Claude Opus~4.5 was able to figure out what GitHub repository the executable originates from, typically by inferring from \texttt{./executable -h} standard output, it then performed a shallow clone of the remote repository to obtain the source code.

Although the task worker still must generate a \texttt{compile.sh} script that builds the executable, the implementation effort, which is typically where the majority of a task worker's efforts are required, becomes trivially simple and yields no insight into the research questions posed by our work.
Another infrequent but still observed failure mode is that a task worker will simply submit a wrapper around the reference executable and call it a day (this is fully mitigated as explained below).

\subsubsection{Early Mitigation Attempts}
\label{appx:benchmark:inference:mitigation}

Before blocking internet access entirely, we tried allowing it while detecting and penalizing cheating after the fact.
This section describes our LM-as-a-judge cheat detection pipeline, its results, and the limitations that led us to disable internet access.

\textit{Detection pipeline.}
We built a rubric-based annotation system that classifies agent trajectories into two violation types: \textit{source code lookup} (cloning the repository, downloading via package managers, reading cached dependency source) and \textit{binary wrapping} (submitting a thin wrapper around the reference executable instead of a real reimplementation).
For each task, 9 LM judges independently review the full command history and classify the trajectory.
To reduce single-model bias, the judges are drawn from three model families: 3 instances of GPT~5.2, 3 of Claude Sonnet~4.5, and 3 of Gemini~3.1 Pro.
A task is flagged if a strict majority (5 or more of 9) judges identify at least one violation.

\textit{Cheating rates.}
We ran this pipeline on internet-access runs for four models: Claude Opus~4.6, Claude Sonnet~4.6, Gemini~3 Flash, and GPT~5~mini.
As reported in Table~\ref{tab:cheating}, cheating rates range from 1\% (GPT~5~mini) to 36\% (Claude Sonnet~4.6).
Source code lookup is the main violation type, accounting for 79--95\% of flagged tags across the three models with meaningful cheating rates.
Models use a range of strategies: directly cloning the GitHub repository, installing the project via a package manager (\texttt{cargo install}, \texttt{go get}, \texttt{apt-get source}), or reading cached dependency source from local package caches such as \texttt{\textasciitilde/.cargo/registry/src/} and Go's module cache.
This last strategy is especially hard to judge, as it blurs the line between reading dependency code and looking up the project's own source (see below).

\textit{Inter-judge agreement.}
Despite using 9 judges from three model families, agreement is moderate at best.
Fleiss' $\kappa$ ranges from 0.16 (GPT~5~mini) to 0.60 (Claude Sonnet~4.6), with a pooled $\kappa$ of 0.57 across all 786 annotated tasks, which falls in the ``moderate agreement'' range.
Judges disagree on 16--57\% of tasks depending on the model, with the highest disagreement rate for Claude Opus~4.6 (57\%), whose cheating strategies tend to be more subtle.

The main source of disagreement is whether reading dependency source code from local package caches counts as a violation.
For example, on the \texttt{handlr} task (a Rust project), Claude Sonnet~4.6 navigated into \texttt{\textasciitilde/.cargo/registry/src/} and read the source of dependencies like \texttt{xdg-mime} and \texttt{clap}.
Five of nine judges flagged this as source code lookup; the other four called it legitimate, reasoning that these are third-party libraries (not the project itself) that happened to be locally available.

Another gray area is consulting API documentation.
On the \texttt{codesnap} task, Claude Sonnet~4.6 used \texttt{curl} to fetch pages from \texttt{docs.rs} (the Rust documentation hosting service) for the project's published crate.
Four judges flagged this as source lookup; five called it clean, arguing that reading public API docs is closer to consulting a reference manual than obtaining source code.
The task was not flagged (4 of 9 is below the majority threshold), but the split shows how reasonable judges can disagree on where to draw the line.

These disagreements reflect real ambiguity in what counts as ``cheating'' when models have access to package ecosystems and documentation.
Stricter rubrics risk penalizing legitimate reverse engineering strategies; looser ones risk missing real violations.

\begin{figure}[ht]
    \centering
    \begin{minipage}[t]{0.32\textwidth}
        \verbpromptboxfile[PromptBlue]{tables/cheat_disagree_action.txt}
    \end{minipage}%
    \hfill
    \begin{minipage}[t]{0.32\textwidth}
        \verbpromptboxfile[PromptRed]{tables/cheat_disagree_flag.txt}
    \end{minipage}%
    \hfill
    \begin{minipage}[t]{0.32\textwidth}
        \verbpromptboxfile[PromptGreen]{tables/cheat_disagree_clean.txt}
    \end{minipage}
    \caption{Example of a 5--4 judge disagreement. Claude Sonnet~4.6 read dependency source code from the local Cargo registry cache while working on the \texttt{handlr} task. Five judges flagged this as source code lookup (red); four called it legitimate, noting the files are third-party libraries, not the project itself (green).}
    \label{fig:cheat_disagreement}
\end{figure}

\begin{figure}[ht]
    \centering
    \includegraphics[width=\textwidth]{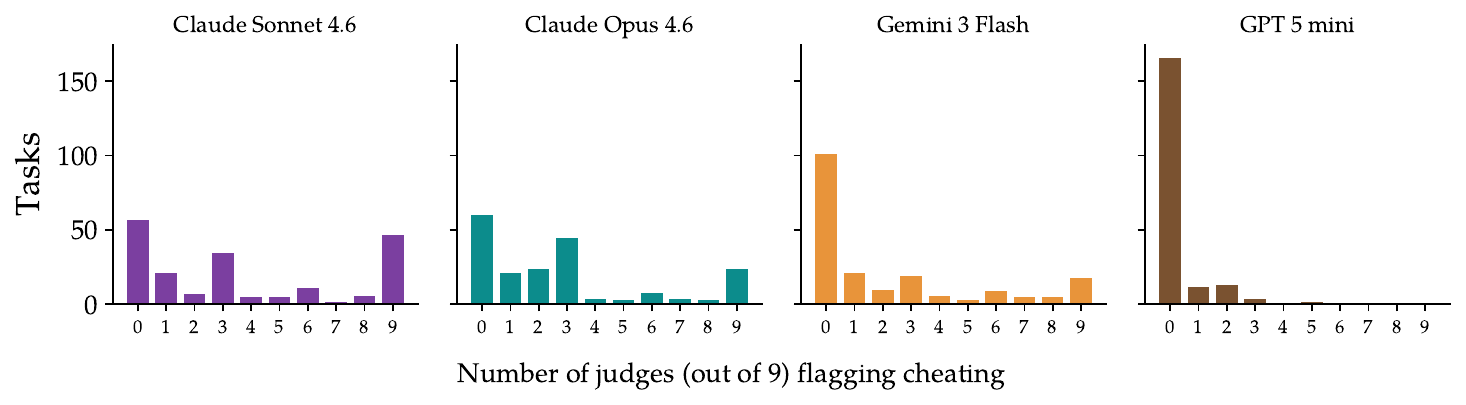}
    \caption{Distribution of judge votes per task for each model evaluated with internet access. Each bar shows how many tasks received exactly $k$ cheating flags from the 9-judge panel. Tasks at $k \geq 5$ are flagged by majority vote. Claude Sonnet~4.6 shows a bimodal pattern (many tasks at 0 and 9), while Claude Opus~4.6 has more mass in the ambiguous middle range, explaining its higher disagreement rate.}
    \label{fig:cheating_vote_distribution}
\end{figure}

\subsubsection{Guidelines}
\label{appx:benchmark:inference:guidelines}
To address these concerns, we attempt to mitigate potential loopholes by imposing several measures and restrictions on the image construction, inference, and evaluation pipelines.

For the Docker image construction pipeline, based on what we discussed in \S\ref{sec:benchmark:construction}, we highlight several key steps aimed at making illegitimate solution patterns such as decompilation and re         discovering local copies of source code impossible or more difficult to achieve:

\begin{itemize}
    \item \textbf{Execute-only permissions}: To prevent task workers from using a tool like \texttt{ghidra} to reverse engineer an executable into code, we set the permissions of the executable to \texttt{111} (execute-only).
    \item \textbf{Removal of original executables during evaluation}: The model might attempt to submit the original executable together with a thin wrapper around it as its solution. This is mitigated by removing all files that match the hash of the original executable prior to evaluation.
    \item \textbf{Inject the executable instead of building from source}: Early on, we used a simpler workflow to construct the Docker image for the starting state of a task.
    Starting with the repository, we would first build the executable by running the model generated \texttt{build.sh} script, then invoke a model generated \texttt{clean.sh} script that would wipe the source code and any artifacts that give away implementation details.
    We found that although models' \texttt{clean.sh} scripts were quite capable of wiping out source code, they would often miss build artifacts and hidden files.
    For instance, at one point we added explicit reminders to remove caches (e.g., \texttt{$\sim$/.cargo/}, \texttt{$\sim$/.cache/**}) and build folders (e.g., \texttt{target/}, \texttt{build/}), which contain direct or suggestive information about the implementation.
    Furthermore, a handful of repositories, typically implemented in C/C++ where build systems are much more heterogenous unlike the Rust or Golang developer ecosystems, would produce artifacts in a much less predictable, uniform manner.
    Consequently, the model would frequently fail to explore and remove such artifacts when generating \texttt{clean.sh}.
    The artifact would then be discovered at inference time.
    Therefore, to eliminate any possibility of such leakage, we opted for the final approach of separating execution compilation and model inference into two separate containers.
    \item \textbf{\texttt{git} history removal}: The original repository's version control history can reveal the remote URL, branch names, and commit messages that identify the project. During image construction, we wipe the existing \texttt{.git/} directory and reinitialize a fresh repository with a single anonymous commit containing only the documentation and the executable.
\end{itemize}

Collectively, these steps ensure that decompilation is impossible, submissions that wrap the gold executable are illegitimate, and local build artifacts that partially or fully give away implementation details do not exist.
We review the restrictions and checks we impose on the inference and evaluation procedures.

\begin{itemize}
    \item \textbf{No Internet Access}: Early in the project, we experimented with several iterations where we allowed internet access at inference time, but then prohibited specific behaviors that we deemed cheating by describing them specifically in the task instructions.
    We also ran an LM-as-a-judge pipeline that would flag cheating by checking the agent trajectory corresponding to a solution.
    Such measures quickly devolved into a cat-and-mouse game, with more capable models coming up with measures to circumvent instructions, such as downloading source code by importing via a dependency manager like \texttt{cargo} rather than directly from Github.
    Furthermore, verbalizing the fine line of what is (not) permitted also quickly became tricky.
    On occasion, models' thought traces exhibit uncertainty over whether certain actions were permissible.
    When working on \texttt{FFmpeg/FFmpeg}, Claude Sonnet 4.6 expressed hesitation over whether it was allowed to download a dependency that happened to be co-located (implemented in the same GitHub repository) with the \texttt{FFmpeg} source code.
    We considered running the LM-as-a-judge cheat detection with multiple models as a way to address ``gray areas'', but found that there were high rates of disagreements and volatility in the judgments themselves.
    Therefore, in favor of simplicity and rigor, we block internet access during \bench{} inference.
    \item \textbf{System prompt instructions}: As shown below, the task instructions inform what behaviors are disallowed.
    Even though the aforementioned guardrails ensure that disallowed actions are futile, we still include this text because without it, task workers may waste turns generating unpermitted actions.
    The full system prompt is shown below.
\end{itemize}

\verbpromptboxfile{assets/system_prompt.txt}

\subsubsection{On the Feasability of ProgramBench}
\label{appx:benchmark:inference:impossible}

The difficulty of the \bench{} task paired with the constraints we impose at inference time may leave some readers wondering whether some, or for that matter, any \bench{} task instances are even possible to resolve.
\bench{} is a very challenging benchmark by today's standards.
While we do not carry out any formal human studies, based on the development history, codebase size, and range of functionality, we venture that an average \bench{} task instance could take an individual or team days, if not weeks' or months' worth of time to complete.

That said, we argue that while formidable, \bench{} task instances are certainly solvable by construction.
The key reason \bench{} is not truly impossible: every test asserts executable behavior that is observable and deterministic.
The same executable is fully accessible to the model, with the necessary permissions to run with any inputs and see exact outputs.
Across the next several paragraphs, we pose and address potential concerns about \bench{}'s resolvability.

\textbf{Can functionality written in one language be reproduced with another?} It is possible that models chooses to implement their solution in different language than one the original source code is written in, which raises the question: does a solution in an alternative programming language even exist?
Computer science theory says yes.
By the Church-Turing thesis~\citep{turing1936computable}, any deterministic input-output behavior that can be realized by a program implemented in one Turing-complete language can be realized by a program written in another Turing-complete language.
Simply put, all general-purpose programming languages are equivalent in computational power.
All languages represented across \bench{} task instances are Turing-complete.
Models are also provided multiple Turing-complete languages in the task environment.

\textbf{Could obscure program behaviors be impossible or arbitrarily hard to discover?} A subtler concern is discoverability of tested behaviors.
To elaborate, given that the test generator runs with full access to the source code, it is possible that some generated tests target hard-to-detect, edge-case behavior.
However, this concern is a reflection of \bench{}'s difficulty, not feasibility.
Models could very well not think of running the executable with certain permutations of flags and inputs.
This does not mean the task is impossible.
Rather, it demonstrates how exploration of program behavior is a key challenge posed by \bench{}, and that beyond writing code, models are tested on their thoroughness when it comes to uncovering behaviors and considering boundary conditions that human developers accounted for.

Conceptually, one scenario where behavior is borderline impossible to discover is if an executable supports functionality that is not communicated or documented via any observable channel.
In other words, there is functionality that is not revealed by the \texttt{README.md}, \texttt{--help} flag standard output, or any artifacts that could be unveiled by typical exploratory actions.
A corresponding test would then effectively be penalizing models for failing to discover behavior that was never observable in the first place.
Such tests may also incentivize models to exhaustively and unintelligibly probe an executable with random inputs, rather than engage in systematic, hypothesis-driven exploration.

In practice, we find that well-maintained programs document their interfaces thoroughly.
Flags, subcommands, supported inputs, and example invocations are frequently surfaced as help output, man pages, or usage documentation.
The above scenario, an executable with important but entirely undiscoverable functionality, would be considered not only bad practice, but even defective by conventional software engineering practices.
Out of caution, we reviewed all 200 repositories for such discrepancies and did not find any instances where an executable's supported behavior was entirely absent from the task instance's start state.


\textbf{What if tasks require internet to solve?}
While creating task instances, we came across several repositories that require internet access for a variety of reasons.
The following are a handful of GitHub repositories that successfully construct an executable, but were not usable as \bench{} task instances because the executable inherently requires communication with one or more endpoints on the web.

\begin{itemize}
    \item \texttt{thomas-mauran/chess-tui}: A terminal chess client. Its core gameplay requires connecting to lichess.org to match with remote opponents.
    \item \texttt{aquaproj/aqua}: A CLI tool manager. It downloads tool binaries from GitHub releases, making the entire tool lifecycle network-dependent.
    \item \texttt{builditluc/wiki-tui}: A terminal Wikipedia reader. Every user interaction fetches and renders articles from the Wikipedia API.
\end{itemize}

We take care to only admit task instances to the test set that do not have this need.
That said, \bench{} nonetheless contains networking tools; 18 task instances are utilities such as HTTP clients, DNS resolvers, and port scanners.
While internet access is cut off, \bench{} task environments still retain loopback networking (\texttt{127.0.0.1}).
The reverse engineering challenge in these tools lies in their protocol handling, output formatting, and CLI logic, not in reaching a remote host.
Therefore, task workers can still develop faithful implementations by spinning up local servers and exercising program behavior against localhost.
We mark the 18 task instances that involve networking functionality in Table~\ref{tab:repository_list}.

\textbf{Do task workers have access to evaluation assets?}
Some behavioral tests exercise the executable with input files such as images, audio files, videos, spreadsheets, or domain specific configurations.
Coupled with the lack of internet access, this highlights an unfair asymmetry where evaluation uses assets that either models are unable to generate on their own or are relatively obscure, executable-specific file formats.

To address this, for each task instance, after generating tests, we run a script to extract such assets.
The script abides by a simple heuristic: keep any files with extensions that aren't on a ``blocklist'' of popular, standardized, text-based file formats.
The rationale is that we are not interested in the failure mode where models fail to develop a solution because it could not come up with characteristic test data.
This means that binaries (e.g., \texttt{png}, \texttt{mp3}, \texttt{wav}, \texttt{xlsx}) are always provided, as today's models may not be capable of synthesizing such files directly.
A handful of repositories propose and use their own file formats (e.g., \texttt{.hcl} for HashiCorp configurations); our script also keeps these as well.
On the other hand, files represented in popular formats are not provided; for example, we do not give \texttt{.php} or \texttt{.c} files used by evaluation suites for the \texttt{php/php-src} and \texttt{tinycc/tinycc} task instances respectively.
Just as a human developer could come up with sample files, we impose the same expectations on the task worker.
Coming up with a diverse range of inputs that accounts for edge cases is part of tackling a \bench{} task instance.

As a final note, we have observed evidence of models being able to programmatically generate binary assets.
During test generation for \texttt{FFmpeg/FFmpeg}, models used ffmpeg's built-in \texttt{lavfi} virtual input device to synthesize audio and video on the fly (e.g., \texttt{sine} for waveforms, \texttt{testsrc} for video patterns), avoiding the need for pre-existing media files entirely.
Similarly, during inference across several image-processing tasks (e.g., \texttt{cslarsen/jp2a}, \texttt{hpjansson/chafa}), models wrote inline Python scripts using Pillow to programmatically create test images.
This suggests that going forwards, the need to provide binary assets may diminish as models become more adept at generating their own.

\subsection{Test Generation}
\label{appx:benchmark:test_generation}
We briefly extend upon our motivation for designing novel test generation pipelines, then explore several strategies for automated test generation via \texttt{mini-SWE-agent}.

\textbf{Limitations of skeleton-based evaluations.}
As alluded to in the related works (\S\ref{sec:related}), the preceding common approach to evaluating language models on their ability to build projects to scratch has typically abided by a fill-in-the-blank structure.
This is a direct consequence of the collection and evaluation methodology.
To create a task instance, typically from an open source GitHub repository, class and method implementations are first deleted procedurally using Python-specific libraries like \texttt{ast}.
The repository's original unit test suite is then used for evaluation.
Consequently, if a model deviates from the expected signatures, even if the deviation is reasonable, the test harness won't locate or execute that code.

It is effectively impossible to evaluate truly free-form solutions if tests assert against implementation.
From manual inspection of repositories during early stages of the project, we also found that behavioral checks, such as end-to-end, integration testing, do not exist in the wild anywhere near as frequently as unit testing.
The yield rate of filtering for repositories that not only satisfy \bench{}'s collection constraints, but also have high coverage, end-to-end testing suites is extremely low based on our initial attempts.

\textbf{Frequency of existing testing.}
To motivate the need to generate behavioral tests, we first check to what extent repositories already have such testing.
As mentioned in \S\ref{sec:benchmark:construction}, \bench{} does \textit{not} require repositories have pre-existing tests, nor does the test generation strictly depend on such information.
That said, we perform this investigation to gauge the necessity of our test generation strategies.

Out of all 200 task instances, 141 (70.5\%) have tests, while 59 (29.5\%) do not have an existing test suite.
We detect tests via filename conventions (e.g., \texttt{test\_*.py}, \texttt{*\_test.go}) and directory names (e.g., \texttt{tests/}, \texttt{\_\_tests\_\_/}); a repository is marked as having tests if at least one such file exists.
We find that either the large majority or entirety of existing test suites consist of implementation focused unit tests targeting code-level correctness.
Behavioral test suites that invoke an executable in multiple ways occasionally appear in the codebase, but not nearly consistently enough to obviate the need for generating additional tests.

\subsubsection{Strategies}
\label{appx:benchmark:test_generation:strategies}

To generate tests, we generally first give one or more prompts to a SWE-agent.
The agent is then asked to write the behavioral tests. 
We summarize the approaches we investigate:

\begin{itemize}
    \item \emph{Monolithic}: We give the SWE-agent a single prompt, asking it to generate a comprehensive behavioral test suite in one pass.
    \item \emph{Decomposed}: We give the SWE-agent six specialized prompts, each targeting a narrow category: argument parsing, configuration, help output, I/O behavior, subcommand dispatch, and TUI interaction. The hypothesis is that more variegated prompting leads to more diverse testing.

    \item \emph{Coverage-Guided Iterative}: We use the SWE-agent to explore the program, its source code, existing tests, and documentation, and then generate behavioral tests. We also prompt the agent to identify and include in its test suite any existing behavioral tests defined in the repository (\textit{harvesting}). The agent continuously measures the line coverage of the current test suite and iteratively writes new tests to invoke missing code paths, attempting to achieve full coverage.

    Naturally, some generated tests may have missing or trivially true assertions. Therefore, to ensure assertion quality, tests are flagged if they fail the gold binary or trigger our assertion quality linter, which detects structurally weak assertion patterns such as exit-code-only checks, short substring matches, and disjunctive assertions (see \ref{app:lint-rules} for full list of rules). The agent is prompted to revise all flagged tests. The loop continues until the suite satisfies a target coverage threshold.
\end{itemize}

At the end of every strategy, we discard any tests that do not pass with the gold binary deterministically or pass a dummy binary.

\begin{table}[b]
    \centering
    \small
    \caption{The six specialized configs used in the \emph{decomposed} test generation strategy. Each config prompts a SWE-agent to write tests targeting a narrow behavioral category of the executable.}
    \label{tab:decomposed_configs}
    \begin{tabular}{@{}ll@{}}
        \toprule
        \textbf{Type} & \textbf{Description} \\
        \midrule
        Args        & Flag formats, required/optional arguments, positional args, type validation \\
        Config      & Environment variables, config file loading, precedence rules \\
        Help        & \texttt{--help}/\texttt{-h} output, usage synopsis, subcommand help text \\
        I/O         & Stdin/file input, stdout/stderr separation, exit codes, output formatting \\
        Subcommand  & Subcommand dispatch, routing, global vs.\ local flags, aliases \\
        TUI         & Interactive navigation, key bindings, screen state, mode switching \\
        \bottomrule
    \end{tabular}
\end{table}


\subsubsection{Analyses}
\label{appx:benchmark:test_generation:analyses}

We provide several statistical and qualitative breakdowns of the proposed test generation strategies.

\begin{figure*}[t]
    \centering
    \includegraphics[width=\textwidth]{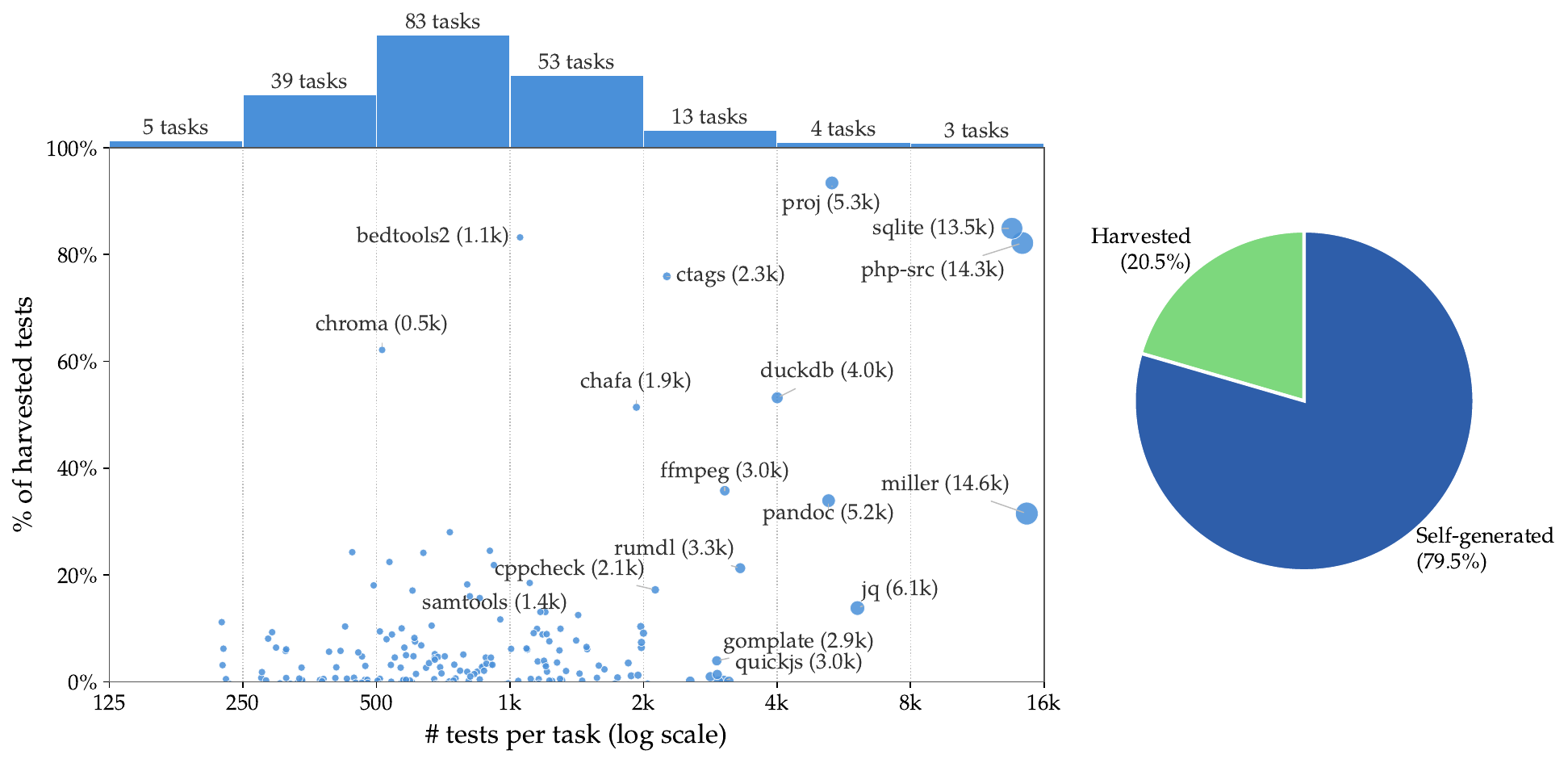}
    \caption{Distribution of test volume and source across all \bench{} task instances. Left: each dot is one task; the x-axis shows the total number of tests (log scale) and the y-axis shows the percentage of tests harvested from existing repository test suites versus self-generated tests. Bin counts are annotated above the plot. Right: global split of harvested vs.\ self-generated tests across all tasks.}
    \label{fig:harvest_scatter}
\end{figure*}

\textbf{The coverage-guided iterative strategy does the best.}  The monolithic strategy averages 27.8 tests per task. The decomposed strategy does better, averaging 51.7 tests per task. However, the coverage-guided iterative strategy results in a significantly larger number of tests with a median of 750 tests per task and with most tasks having between 200 to 2000 tests. It also results in a very high mean line coverage of 79.7\%, with a median of 86.2\% (Figure~\ref{fig:coverage-scatter}). Furthermore, when possible, the agent effectively makes use of existing tests, e.g., SQLite, PHP, Bedtools2, Proj and Ctags (Figure~\ref{fig:harvest_scatter}).

\textbf{The majority of tests are generated, not harvested.}
Our test generation pipeline combines two sources: tests generated from scratch via the strategies described above, and tests harvested from existing repository test suites that exercise executable behavior, as done in the `coverage-guided iterative' strategy.
Figure~\ref{fig:harvest_scatter} shows the distribution of total tests per task, and breaks down the proportion from each source.
The majority of tests are self-generated (79.5\%), while the remaining 20.5\% are harvested from existing suites.

\textbf{Incorporating assertion-quality signals generates stronger tests.} A static linter plus gold/dummy execution feedback in the test generation loop reduces the mean dummy pass rate from 18.5\% to 3.7\% (a 5$\times$ reduction) and eliminates the worst-case failure mode where a large fraction of a task's tests are trivially passable. On the same set of tasks evaluated across four frontier models (Opus 4.6, Sonnet 4.6, Gemini 3.1 Pro and GPT-5.4), the resulting tests are 20–30 percentage points harder to pass than tests generated without these signals, with no
change in model rank ordering, confirming the gain comes from stricter assertions rather than from incidentally harder tests.

\textbf{Coverage measurement methodology.}
For each task, we merge all active generated test branches and execute them against a coverage-instrumented build of the original binary.
We measure first-party line coverage only, excluding system headers, vendored dependencies, and auto-generated code.
The coverage scatter plot (Figure~\ref{fig:coverage-scatter}) shows a representative subsample of 100 tasks drawn from the 154 tasks where we independently performed coverage measurement, disregarding the coverage recorded by the test generating agent.

\textbf{Behavioral suite baseline selection.}
From our benchmark's hard tasks (difficulty score $\geq 4$; see \S\ref{appx:benchmark:statistics} for the scoring formula), we selected twelve repositories that maintain an identifiable application-level behavioral or integration test suite, as opposed to only internal unit tests.
The selected suites are: PHP's \texttt{.phpt} regression harness, FFmpeg's FATE suite, Typst's CLI and rendering tests, Miller's regression cases, StGit's shell workflow tests, Doxygen's documentation fixtures, Lazygit's integration tests, Atlas's CLI/e2e plus SQLite integration tests, Rumdl's integration tree, JSON Schema's CLI plus conformance tests, Zstd's CLI and fuzzer suite, and jq's regression harness.
We audited each baseline against upstream CI configurations to ensure the comparison is integration-test only, excluding \texttt{src/} unit tests in projects (Rumdl, Atlas) where the default test runner would otherwise inflate native coverage with code paths our black-box generated tests structurally cannot reach.

\newpage
\subsubsection{Test Examples}
\label{appx:benchmark:test_generation:examples}

We present examples of tests generated by our pipeline for a select set of task instances.
Generally, all tests are represented in Python using the \texttt{pytest} library, a design that is explicitly requested by our instructions.

\noindent
\begin{minipage}[t]{0.48\textwidth}
\codebox{assets/test_example_zstd_corruption.txt}
\vspace{0.3\baselineskip}
{\small \textbf{zstd} (corruption detection): \texttt{zstd} is a compression tool. The test compresses random bytes (L2), then runs \texttt{-{}-test}, which verifies archive integrity without decompressing (L5). It then flips a single byte in the compressed file (L8--9) and re-runs \texttt{-{}-test}, asserting that the corruption is now detected (L12).}
\end{minipage}
\hfill
\begin{minipage}[t]{0.48\textwidth}
\codebox{assets/test_example_duckdb.txt}
\vspace{0.3\baselineskip}
{\small \textbf{DuckDB} (CSV import and query): \texttt{duckdb} is an analytical database with a CLI shell. The test writes a CSV file to disk (L2--4), then uses the \texttt{.import} dot-command to load it into a table (L6). It immediately queries the table with a \texttt{WHERE} filter (L7--8) and asserts that only the matching row appears in the output (L11--12).}
\end{minipage}

\vspace{0.8\baselineskip}

\noindent
\begin{minipage}[t]{0.48\textwidth}
\codebox{assets/test_example_php.txt}
\vspace{0.3\baselineskip}
{\small \textbf{PHP} (script execution): \texttt{php} is the PHP language interpreter. The test writes a small PHP script to disk (L2--7) that prints the argument count (\texttt{\$argc}) and a comma-joined argument list (\texttt{\$argv}). It invokes the interpreter with \texttt{-n} (no config file), \texttt{-f} (execute script from path), and \texttt{-{}-} (separates interpreter flags from script arguments) on L9. The assertions verify the argument count and that a space-containing argument (\texttt{"b c"}) is passed through correctly (L12--13).}
\end{minipage}
\hfill
\begin{minipage}[t]{0.48\textwidth}
\codebox{assets/test_example_ffmpeg.txt}
\vspace{0.3\baselineskip}
{\small \textbf{FFmpeg} (stdin piping): \texttt{ffmpeg} is a multimedia processing tool. The test first generates a WAV audio file from a silent source (\texttt{-f lavfi} on L3 selects a virtual input device; L4 synthesizes silence at 8\,kHz). It confirms the output has a valid WAV header (L7). A second invocation on L9 pipes the WAV back through \texttt{ffmpeg} via stdin (\texttt{-i -}) with \texttt{-c copy} (stream-copy, no re-encoding), then asserts SHA-256 equality (L14).}
\end{minipage}

\subsubsection{On Test Overspecification}
\label{appx:benchmark:test_generation:over_specification}

As referenced in \S\ref{sec:benchmark:construction}, here we provide an extended discussion about test over-specification and its impact on the feasibility of accomplishing a \bench{} task instance.

Overspecified tests are undesirable because they make a \bench{} task instance infeasible to successfully complete.
Generally, an ``overspecified" test checks a solution for details that are not stated explicitly in the task instructions \textit{and} impossible to discover during the task solving process.
In other words, an overspecified test asserts a condition that is completely inaccessible to a task worker.

\bench{} addresses this issue by virtue of how tests are constructed.
In lieu of unit tests, which we define as tests that check against source code, \bench{} tests assert only on observable behavior of an executable, such as standard output/error, exit codes, and file system modifications.
Therefore, because tests always involve invocations of an executable, they inherently are incapable of targeting details such as variable names or method definitions.

A subtler concern is whether tests might demand exact reproduction of implementation-dependent outputs, such as floating point precision or hash iteration order.
A concurrent case study highlights this tension: MirrorCode~\citep{adamczewski2026mirrorcode} addresses this explicitly by manually vetting 4 repositories ``that seemed feasible for a human software engineer to reimplement under similar constraints'', where feasibility is defined as programs where the behavior and documentation are closer to a specification rather than a reverse engineering challenge.

We take a different approach.
Rather than deciding manually which behaviors are reasonable, we treat the gold executable as the complete specification, which means any deterministic, observable behavior is fair game.
This is a deliberate design choice.
First, the model has full access to the executable at inference time, so any behavior a test checks can be discovered by running the same command.
The suite contains no hidden information that the model cannot obtain itself.
Second, after several rounds of annotations, we came to the conclusion that distinguishing ``meaningful behavior'' from ``implementation artifact'' can be very subjective and does not scale.
Whether the 8th decimal digit of a computation matters depends entirely on the application; for a library used to calculate the trajectory of a satellite in space, precision is of utmost importance.
Third, our gold evaluation procedure filters nondeterministic tests by running each test against the reference executable multiple times and throwing out any that do not pass consistently.
The tests that remain reliably characterize an executable's behavior, which is precisely what we ask the model to reproduce.

We audited \bench{} task instances to investigate whether any task instances are ``at risk'' of overspecification.
We find that in our setting, there are very limited manners in which overspecification can occur; namely, floating point precision, hash/map iteration order, and rendering discretization.
Other potential forms of non-determinism such as timestamp formatting or locale-dependent sorting are really an attribute of \textit{environment} differences, not implementation differences, which our Docker based set up standardizes.
Out of 200 instances, we found only 5 instances where such behavior could plausibly appear in test output:

\begin{itemize}
    \item \texttt{oppiliappan/eva}: a terminal calculator REPL, where arithmetic results could differ in precision depending on the floating point library or operation ordering used by a reimplementation.
    \item \texttt{gromacs/gromacs}: a molecular dynamics simulation toolkit, where simulation outputs involve extensive floating point computation sensitive to operation ordering.
    \item \texttt{rs/jplot}: a terminal plotting tool, where mapping continuous values to discrete character-grid positions involves rounding decisions (e.g., floor vs.\ round) that affect visual output.
    \item \texttt{OSGeo/PROJ}: a cartographic projection library, where coordinate transformations involve chained floating point operations and intermediate rounding.
    \item \texttt{OSGeo/gdal}: a geospatial data translator, where raster/vector transformations and coordinate reprojections involve floating point arithmetic sensitive to operation ordering.
\end{itemize}

We manually inspected the test suites for these five instances and did not find evidence of assertions on implementation-dependent numerical output; the tests predominantly check CLI flag behavior, file format handling, and string output.

\subsubsection{Assertion Lint Rules}
\label{app:lint-rules}

Table~\ref{tab:lint-rules} lists all rules checked by the assertion linter.

\begin{table}[t]
\centering
\small
\begin{tabular}{llp{7.5cm}}
\toprule
\textbf{Rule} & \textbf{Sev.} & \textbf{Description} \\
\midrule
\texttt{no\_assertions}               & HIGH & Test contains no \texttt{assert} statements \\
\texttt{trivially\_true}              & HIGH & \texttt{assert True} or \texttt{assert X or True} \\
\texttt{sole\_returncode}             & HIGH & Only assertion checks \texttt{returncode == 0} \\
\texttt{returncode\_in\_list}         & HIGH & \texttt{assert returncode in [...]} launders non-zero exits \\
\texttt{pass\_body}                   & HIGH & Test body is just \texttt{pass} \\
\texttt{assertion\_disjunction}       & HIGH & \texttt{assert A or B}: should assert exact outcome \\
\texttt{if\_no\_else}                 & HIGH & \texttt{if} branch asserts with no \texttt{else}: silently passes when condition is false \\
\texttt{if\_else\_both\_assert}       & HIGH & Both \texttt{if}/\texttt{else} branches assert: suggests non-deterministic behavior \\
\texttt{try\_except\_swallow}         & HIGH & \texttt{except} handler with \texttt{pass} swallows failures \\
\texttt{all\_assertions\_weak}        & HIGH & All assertions only check \texttt{returncode}, \texttt{len}, or \texttt{isdigit} \\
\texttt{short\_substring}             & HIGH & Substring check shorter than 15 characters \\
\texttt{golden\_written\_in\_test}    & HIGH & Test writes to its own golden file (always passes) \\
\texttt{golden\_no\_equality}         & HIGH & Golden file referenced in docstring but never compared with \texttt{==} \\
\texttt{golden\_docstring}            & HIGH & Golden file mentioned in docstring not found in test body \\
\midrule
\texttt{for\_no\_guard}               & MED  & Loop-only assertions with no pre-loop length check \\
\texttt{weak\_sole\_assertion}        & MED  & Sole assertion is \texttt{len(x) > N} \\
\texttt{relative\_length\_assertion}  & MED  & \texttt{assert len(x) >= N}: relative bound verifies nothing about content \\
\texttt{any\_all\_no\_guard}          & MED  & \texttt{any()}/\texttt{all()} with no non-empty guard \\
\texttt{file\_exists\_no\_content}    & MED  & \texttt{path.exists()} with no content assertion \\
\texttt{only\_negative\_assertions}   & MED  & All assertions are negative (\texttt{not in}, \texttt{!=}) \\
\midrule
\texttt{catches}                      & LOW  & Missing or too-short \texttt{CATCHES:} docstring \\
\bottomrule
\end{tabular}
\caption{Assertion lint rules. HIGH rules indicate tests likely to pass trivially incorrect implementations. MED rules indicate weak but not necessarily vacuous assertions.}
\label{tab:lint-rules}
\end{table}

\textbf{Examples of Weak Assertions.}
The following illustrates common failure modes our linter detects, drawn from real generated tests.

\noindent
\begin{minipage}[ht]{0.48\textwidth}
\codebox{assets/lint_example_exitcode.txt}
\vspace{0.3\baselineskip}
{\small Exit-code-only assertion: passes any implementation that runs without crashing. Ignores the contents of the help text.}
\end{minipage}
\hfill
\begin{minipage}[ht]{0.48\textwidth}
\codebox{assets/lint_example_substring.txt}
\vspace{0.3\baselineskip}
{\small Short substring check: the 3-character literal \texttt{"*"} matches trivially.}
\end{minipage}

\vspace{0.8\baselineskip}

\noindent
\begin{minipage}[ht]{0.48\textwidth}
\codebox{assets/lint_example_disjunction.txt}
\vspace{0.3\baselineskip}
{\small Disjunctive assertion: accepts either outcome rather than asserting the true expected output.}
\end{minipage}
\hfill
\begin{minipage}[ht]{0.48\textwidth}
\codebox{assets/lint_example_tryexcept.txt}
\vspace{0.3\baselineskip}
{\small \texttt{try/except} swallowing failures: any exception silently passes. Also only asserts on content length, not content itself.}
\end{minipage}

\subsection{Dataset Statistics}
\label{appx:benchmark:statistics}

In the following sections, we provide breakdowns of the \bench{} dataset by various dimensions, including programming languages, repository size, and file extensions.
These statistics characterize the diversity of the \bench{} dataset and can help inform future analyses and evaluations.
In general, our typical approach to these analyses is to perform simple, straightforward aggregations (e.g., line, file counts) or keyword-based searches (e.g., \texttt{test\_*.go} to identify test files), unless specified otherwise explicitly.
GitHub community statistics (stars, contributors, commits) were crawled on April 21, 2026.

\begin{figure}[h!]
    \centering
    \begin{minipage}[t]{0.48\textwidth}
        \centering
        \includegraphics[width=\textwidth]{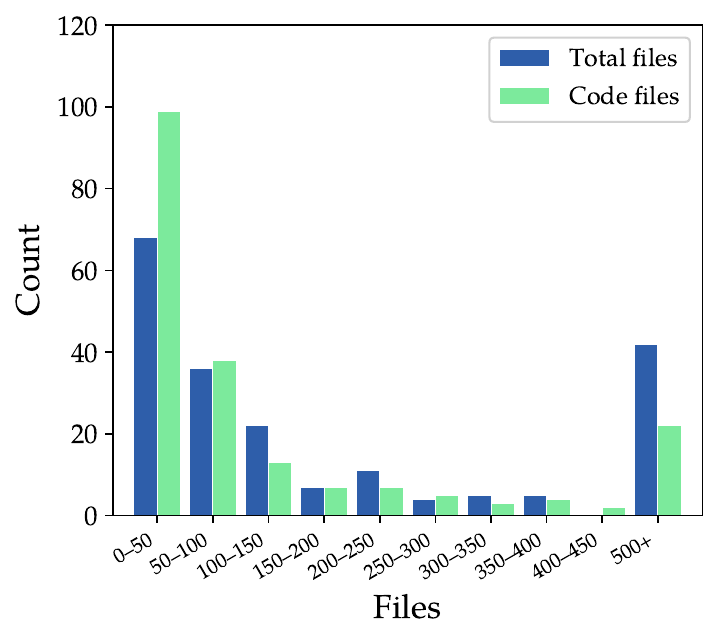}
        \captionof{figure}{Distribution of total/code files per repository.}
        \label{fig:files_overlaid}
    \end{minipage}
    \hfill
    \begin{minipage}[t]{0.48\textwidth}
        \centering
        \includegraphics[width=\textwidth]{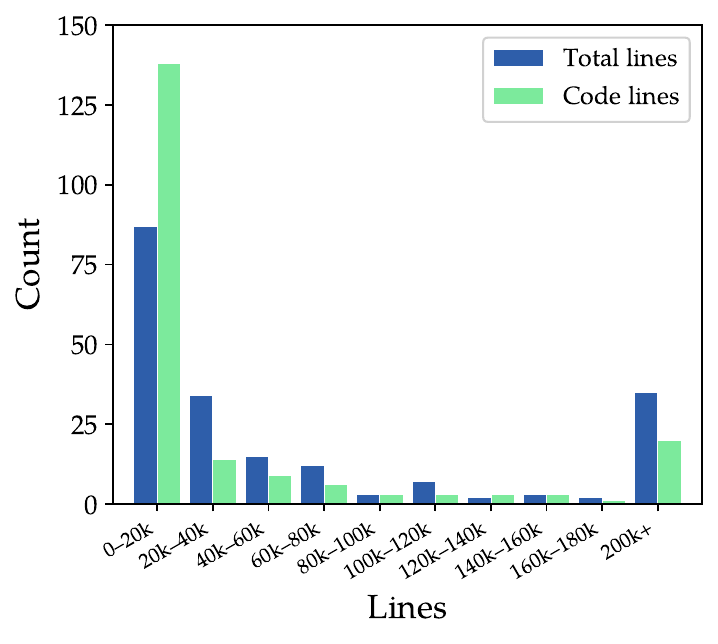}
        \captionof{figure}{Distribution of total/code lines per repository.}
        \label{fig:lines_overlaid}
    \end{minipage}

    \vspace{0.5\baselineskip}

    \begin{minipage}[t]{0.42\textwidth}
        \centering
        \includegraphics[width=\textwidth]{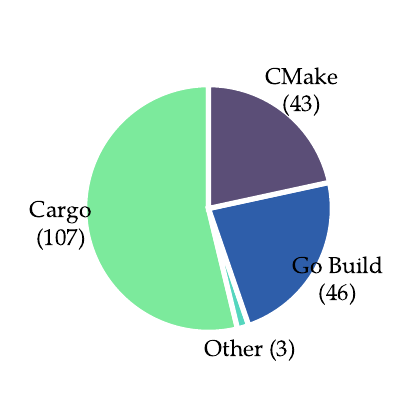}
        \captionof{figure}{Distribution of build systems across \bench{} repositories. The build system distribution closely mirrors the language distribution (Figure~\ref{fig:dist_language}), as each language ecosystem has a dominant build tool.}
        \label{fig:dist_build_system}
    \end{minipage}
    \hfill
    \begin{minipage}[t]{0.52\textwidth}
        \centering
        \includegraphics[width=\textwidth]{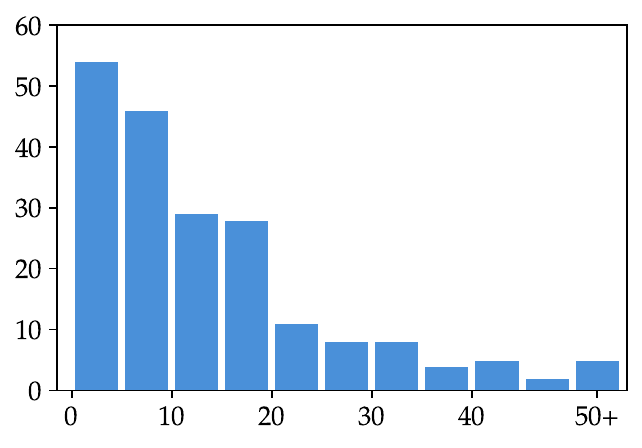}
        \captionof{figure}{Distribution of runtime dependencies across \bench{} repositories. Counts are extracted by parsing root-level package manifests (\texttt{Cargo.toml}, \texttt{go.mod}, \texttt{CMakeLists.txt}, etc.), excluding development and test dependencies.}
        \label{fig:dist_dependency_count}
    \end{minipage}
\end{figure}

\textbf{Number of files and lines.} Figures~\ref{fig:files_overlaid} and \ref{fig:lines_overlaid} show the distribution of repository sizes in \bench{} by number of files and lines.
The median codebase has 93 total files with 50 code files, and 8,635 lines of code spread across all files.
The extremes of the \bench{} dataset feature several prodigious codebases.
\texttt{FFmpeg/FFmpeg} has the most code files at 4566 total, while \texttt{php/php-src} (the official source code for the PHP interpreter) is implemented with 1.97 million lines of code.
In comparison to existing benchmarks, these statistics confirm how \bench{} is a step function increase in terms of implementation scale.

\begin{figure}[t]
    \centering
    \includegraphics[width=\textwidth]{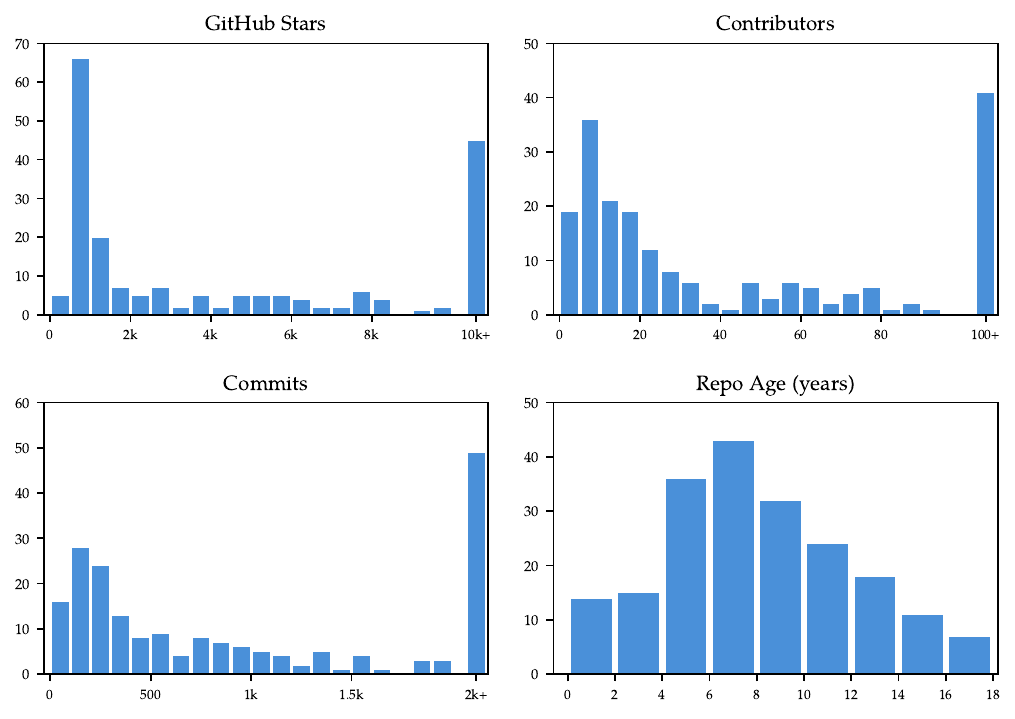}
    \caption{Distribution of GitHub community and development history metrics across \bench{} repositories: stars, contributors, commits, and repository age.}
    \label{fig:contribution_stats}
\end{figure}

\textbf{Build systems.} A repository's primary language is determined by counting lines of code per language (mapped from file extensions) and selecting the one with the most lines (see Figure~\ref{fig:dist_language} in the main paper).
From Figure~\ref{fig:dist_build_system}, the distribution of build systems used correspond with these ratios.
Build systems are identified by scanning for sentinel marker files (e.g., \texttt{Cargo.toml}, \texttt{go.mod}, \texttt{CMakeLists.txt}), with the primary build system being the one whose marker appears at the repository root.
For Rust and Golang repositories, the \texttt{cargo} and native \texttt{go build} constitutes the entirety of build systems with which the corresponding executables are compiled respectively.
There is relatively slightly more variance with C/C++ repositories due to the differences in libraries that certain projects rely on, but the build procedure generally remains quite uniform.

\textbf{Directory depth.} Repositories in \bench{} are generally shallow: the median maximum depth is 3, with an average file depth of 2.5 levels from the repository root, and a median of 13 directories per repository.
We compute depth by measuring each file's distance from the repository root.
While most repositories (73\%) have maximum depth $\leq$ 4, deeper hierarchies exist (up to 13 levels).
Rust repositories tend to be the flattest (median 9 directories), while C/C++ projects are structurally more complex (median 32 directories), reflecting the heavier use of nested module hierarchies and vendored dependencies common in C/C++ ecosystems.
At the extremes, 3 repositories consist of a single flat directory with no nesting at all (e.g., \texttt{seqtk}, \texttt{tty-clock}), while the deepest (\texttt{gromacs/gromacs} at depth 13) contains over 850 directories.

\textbf{Dependency count.} Of the 200 repositories, 171 (85.5\%) contain a recognized package manifest file; among these, the median repository declares 17 total dependencies (12 runtime).
Dependencies are counted by parsing root-level manifest files (\texttt{Cargo.toml}, \texttt{go.mod}, \texttt{package.json}, etc.) and summing declared packages.
This distribution, with 20.0\% of repositories in the moderate range (16 to 30 dependencies) and 11.0\% heavy ($>$30), implies that successful reconstruction often requires correctly identifying and integrating third-party libraries rather than implementing all functionality from scratch.

\textbf{Contribution statistics.} To proxy how much human effort a repository corresponds to, we present some observations of contribution-related metrics on GitHub (Figure~\ref{fig:contribution_stats}).
The statistics generally showcase how the GitHub repositories represented in \bench{} span a wide range of development lifetimes, with no bias towards any particular statistic, aside from the fact that it produces an executable.

Repositories span a wide range of community adoption, from projects like \texttt{HaliteChallenge/Halite} with 202 stars to widely used utilities like \texttt{junegunn/fzf} with over 79,000 stars (median of 2,124 stars).
The distribution is roughly split in half: 49\% of repositories have fewer than 2,000 stars, while 22\% exceed 10,000.
Many popular tools, such as  \texttt{ffmpeg}, \texttt{jq}, \texttt{tinycc}, \texttt{php-src}, \texttt{ripgrep}, and \texttt{zstd} are represented.
Similarly, projects have a wide range of development history, from \texttt{unhappychoice/gittype} created in late 2025, to \texttt{xorg62/tty-clock} which was seeded in 2008.
The majority of repositories (56\%) were created 4 to 10 years ago, with 30\% being older than a decade.
Solo projects with just a single contributor like \texttt{madler/pigz} account for 5 task instances.
Others like \texttt{typst/typst} reflect larger collaborations with over 400 contributors (median of 22 contributors).
Most repositories are small-team projects: 47\% have fewer than 20 contributors, though 21\% have over 100.
Development effort varies dramatically as well, from \texttt{ip7z/7zip} with 13 commits to \texttt{php/php-src} with over 145,000 commits.
About 44\% of repositories have fewer than 500 commits, while 24\% exceed 2,000.

\begin{table}[t]
    \centering
    \small
    \begin{NiceTabular}{llrl}
    \toprule
    \textbf{Category} & \textbf{Subcategory} & \textbf{Count} & \textbf{Representative Examples} \\
    \midrule
    \multirow{3}{*}{CLI Utilities}
    & Text Processing & 31 & {\scriptsize \texttt{wfxr/csview}, \texttt{sclevine/yj}, \texttt{elkowar/pipr}} \\
        & File \& Disk & 26 & {\scriptsize \texttt{canop/broot}, \texttt{sitkevij/hex}, \texttt{jarun/nnn}} \\
        & System \& Network & 22 & {\scriptsize \texttt{rs/curlie}, \texttt{chmln/handlr}, \texttt{htop-dev/htop}} \\
    \midrule
    \multirow{4}{*}{Developer Tools}
        & Build \& Codegen & 24 & {\scriptsize \texttt{esubaalew/run}, \texttt{pemistahl/grex}, \texttt{cweill/gotests}} \\
        & Linters \& Formatters & 18 & {\scriptsize \texttt{paradigmxyz/solar}, \texttt{mgechev/revive}, \texttt{rvben/rumdl}} \\
        & Documentation & 9 & {\scriptsize \texttt{crowdagger/crowbook}, \texttt{typst/typst}, \texttt{jgm/pandoc}} \\
        & VCS/Git & 9 & {\scriptsize \texttt{jesseduffield/lazygit}, \texttt{foriequal0/git-trim}, \texttt{stacked-git/stgit}} \\
    \midrule
    Terminal Fun \& Demos & --- & 12 & {\scriptsize \texttt{wintermute-cell/ngrrram}, \texttt{jrnxf/thokr}, \texttt{abishekvashok/cmatrix}} \\
    Media \& Graphics & --- & 12 & {\scriptsize \texttt{cslarsen/jp2a}, \texttt{thezoraiz/ascii-image-converter}, \texttt{sharkdp/pastel}} \\
    Productivity \& Notes & --- & 10 & {\scriptsize \texttt{naggie/dstask}, \texttt{lfos/calcurse}, \texttt{cheat/cheat}} \\
    Security \& Forensics & --- & 7 & {\scriptsize \texttt{filosottile/age}, \texttt{rbakbashev/elfcat}, \texttt{sirwart/ripsecrets}} \\
    Database \& Data & --- & 7 & {\scriptsize \texttt{ariga/atlas}, \texttt{multiprocessio/dsq}, \texttt{skeema/skeema}} \\
    Compression \& Encoding & --- & 7 & {\scriptsize \texttt{facebook/zstd}, \texttt{madler/pigz}, \texttt{tukaani-project/xz}} \\
    Languages \& Interpreters & --- & 6 & {\scriptsize \texttt{lua/lua}, \texttt{nuta/nsh}, \texttt{hush-shell/hush}} \\
    \bottomrule
    \end{NiceTabular}
    \caption{Distribution of \bench{} task instances across functional categories. Each repository is assigned to a category based on its primary utility. Representative examples showcase the diversity within each category.}
    \label{table:category_breakdown}
\end{table}

\textbf{Functional categories.} Table~\ref{table:category_breakdown} presents a breakdown of \bench{} task instances by functional category, illustrating the diverse range of software utilities represented in the dataset.
The 200 repositories span 14 distinct categories, with CLI utilities for text processing (31) and file/disk operations (26) being the most prevalent, followed by developer tools for build systems and code generation (24), and system/network utilities (22).
The dataset also includes specialized domains such as security and forensics tools (7), programming language interpreters (6), and compression utilities (7).
This breadth of functional categories ensures that \bench{} tests a task worker's ability to reverse engineer software across a wide spectrum of real-world applications, from everyday productivity tools like \texttt{dstask} and \texttt{calcurse}, to foundational infrastructure like the Lua interpreter and Facebook's \texttt{zstd} compression library.

\textbf{Task difficulty.}
To enable difficulty-stratified analyses, we assign each task a scalar difficulty score on a 0--10 scale derived from two repo-intrinsic metrics: lines of code and number of runtime dependencies.
Concretely, the score is computed as $\text{score} = \text{clamp}\!\bigl(\log_{10}(\text{code\_lines}) + \log_{10}(1 + \text{runtime\_deps}) - 2,\; 0,\; 10\bigr)$, where the $-2$ shift accounts for the baseline produced by even trivial projects and the clamp caps the scale.
Lines of code are counted across files with recognized code extensions (e.g., \texttt{.py}, \texttt{.rs}, \texttt{.go}, \texttt{.c}), while runtime dependencies are extracted from root-level package manifests (\texttt{Cargo.toml}, \texttt{go.mod}, \texttt{CMakeLists.txt}, etc.), excluding development and test dependencies.
We assign fixed, dataset-independent thresholds to bin tasks into three difficulty levels: \emph{Easy} ($< 2$), \emph{Medium} ($2 \leq \text{score} < 4$), and \emph{Hard} ($\geq 4$).
To build intuition for the scale: a small single-file project like \texttt{pingu} (212 lines, 3 dependencies) scores 0.93 and falls squarely in the Easy range.
A moderately-sized project with a handful of libraries (for instance, \texttt{direnv/direnv} at ${\sim}$8k lines and 3 dependencies, for instance) lands around 2.5, typical of the Medium bin.
The Hard designation generally a task has either a very large codebase or a combination of substantial code and many dependencies; \texttt{lazygit}, with ${\sim}$593k lines and 36 dependencies, scores 5.3, while \texttt{FFmpeg} reaches 4.25 from its 1.8M lines alone.

We note that 22 repositories use custom or non-standard build systems (e.g., hand-written \texttt{Makefile}s or bespoke \texttt{./configure} scripts) for which no recognized package manifest was detected, resulting in a runtime dependency count of zero for those tasks; their difficulty scores therefore reflect code size alone.
Table~\ref{tab:difficulty_distribution} summarizes the distribution across bins; per-task scores and labels are listed in Table~\ref{tab:repository_list}.

\begin{table}[t]
    \centering
    \small
    \begin{tabular}{@{}lccl@{}}
        \toprule
        \textbf{Difficulty} & \textbf{Score Range} & \textbf{Count} & \textbf{Examples} \\
        \midrule
        Easy   & $[0, 2)$   & 28  & \texttt{tty-clock}, \texttt{entr}, \texttt{cmatrix}, \texttt{figlet}, \texttt{seqtk} \\
        Medium & $[2, 4)$   & 143 & \texttt{ripgrep}, \texttt{fzf}, \texttt{jq}, \texttt{fd}, \texttt{zstd}, \texttt{htop} \\
        Hard   & $[4, 10]$  & 29  & \texttt{lazygit}, \texttt{FFmpeg}, \texttt{typst}, \texttt{php-src}, \texttt{bat} \\
        \bottomrule
    \end{tabular}
    \caption{Distribution of \bench{} tasks across difficulty bins, with representative repositories.}
    \label{tab:difficulty_distribution}
\end{table}

As a sanity check, we look at whether our difficulty score correlates with observed agent performance.
Table~\ref{tab:difficulty_performance} reports the average test pass rate within each difficulty bin by model.
Pass rates decrease monotonically from \textit{Easy} to \textit{Hard} across all models, suggesting that the metric captures meaningful variation in task complexity.

\begin{table}[t]
    \centering
    \small
    \begin{tabular}{@{}lccc|c@{}}
        \toprule
        \textbf{Model} & \textbf{Easy} & \textbf{Medium} & \textbf{Hard} & \textbf{Avg} \\
        \midrule
        Claude Opus 4.6   & \textbf{73.9\%} & \textbf{53.3\%} & 24.7\% & \textbf{52.0\%} \\
        Claude Opus 4.7   & 73.8\% & 52.1\% & 25.0\% & 51.2\% \\
        Claude Sonnet 4.6 & 67.1\% & 48.8\% & \textbf{28.6\%} & 48.5\% \\
        GPT 5.4           & 50.9\% & 39.9\% & 18.0\% & 38.3\% \\
        Gemini 3.1 Pro    & 50.9\% & 37.7\% & 17.8\% & 36.6\% \\
        Gemini 3 Flash    & 49.6\% & 31.9\% & 17.9\% & 32.4\% \\
        Claude Haiku 4.5  & 41.9\% & 30.6\% & 15.6\% & 30.0\% \\
        GPT 5.4 mini      & 24.1\% & 17.4\% &  7.8\% & 16.9\% \\
        GPT 5 mini        & 21.3\% & 15.9\% & 11.0\% & 15.9\% \\
        \bottomrule
    \end{tabular}
    \caption{Macro-averaged test pass rate by difficulty bin: for each task instance we compute the fraction of tests passed, then average these fractions within each bin. Rates decrease monotonically from Easy to Hard for every model.}
    \label{tab:difficulty_performance}
\end{table}

\newpage
\newpage
\section{Additional Results}
\label{appx:results}

We present additional figures and discussion that complement the experiment setup discussion (\S\ref{sec:experiments}), main evaluation (\S\ref{sec:results}), and analyses (\S\ref{sec:analysis}).

\subsection{Experimental Setup}
\label{appx:results:exp_setup}

Beyond its adoption by widely used leaderboards (\S\ref{sec:experiments}), \texttt{mini-SWE-agent} is built to be neutral in how models interact with a codebase.
Each turn, a model simply issues a bash action which is then executed directly; there are no manually crafted tools or prompts that could unfairly advantage or disadvantage particular models.
This minimal design ensures that performance differences can be attributed directly to model capabilities rather than idiosyncrasies in agentic harnesses.

All configurations below are specified declaratively in \texttt{mini-SWE-agent}'s YAML configuration files, which control the agent loop, container provisioning, and model invocation.

\begin{itemize}
    \item \textbf{Per-action timeout.} Each agent action must complete within 3 minutes; actions exceeding this limit are terminated with a descriptive error message. Models can work around this by launching background processes and polling output files.
    \item \textbf{Cost limit.} No cost limit is imposed. For reported costs, prompt caching is enabled and accounted for across all models.
    \item \textbf{Output truncation.} Command outputs exceeding 10,000 characters are truncated, showing the first and last 5,000 characters with an elision notice.
    \item \textbf{Soft warnings.} When fewer than 20 steps or 10 minutes remain, the agent receives a warning to wrap up and ensure its solution compiles.
\end{itemize}

\subsection{Further Findings}
\label{appx:results:main}

\begin{figure}[ht]
    \centering
    \includegraphics[width=\textwidth]{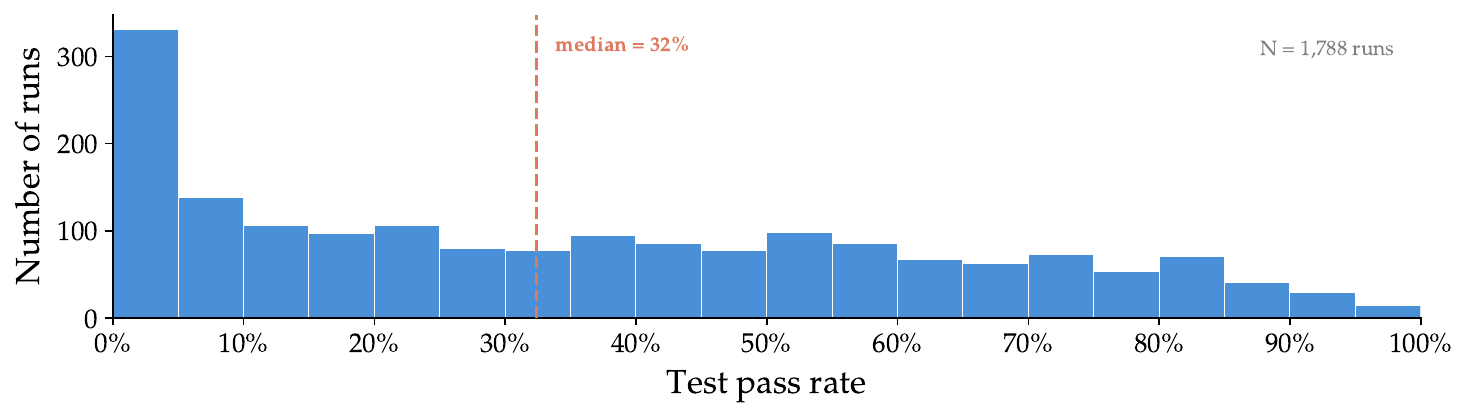}
    \caption{Distribution of test pass rates across all 1,800 leaderboard runs in 5\% bins. The median pass rate across all runs is 32\%.}
    \label{fig:score_histogram}
\end{figure}

\textbf{Most runs achieve meaningful partial progress rather than all-or-nothing outcomes.}
We examine how test pass rates are distributed across all leaderboard runs to understand how partial progress varies across models and task instances.
Figure~\ref{fig:score_histogram} shows the distribution across all 1,788 runs.
The 0--5\% bin is the single largest, containing roughly 18\% of runs, indicating that models do frequently fail to make any meaningful progress on a task.
Beyond that initial spike, however, scores are spread roughly uniformly across the pass-rate spectrum, with a median of 32\%.
The distribution tapers gradually at the high end: fewer than 5\% of runs exceed 90\%, and near-perfect reimplementations are rare.
The large low-scoring bin provides a floor of genuinely difficult tasks, the uniform middle range rewards incremental progress, and the sparse right tail leaves room for future models to improve.

\begin{figure}[ht]
    \centering
    \includegraphics[width=\textwidth]{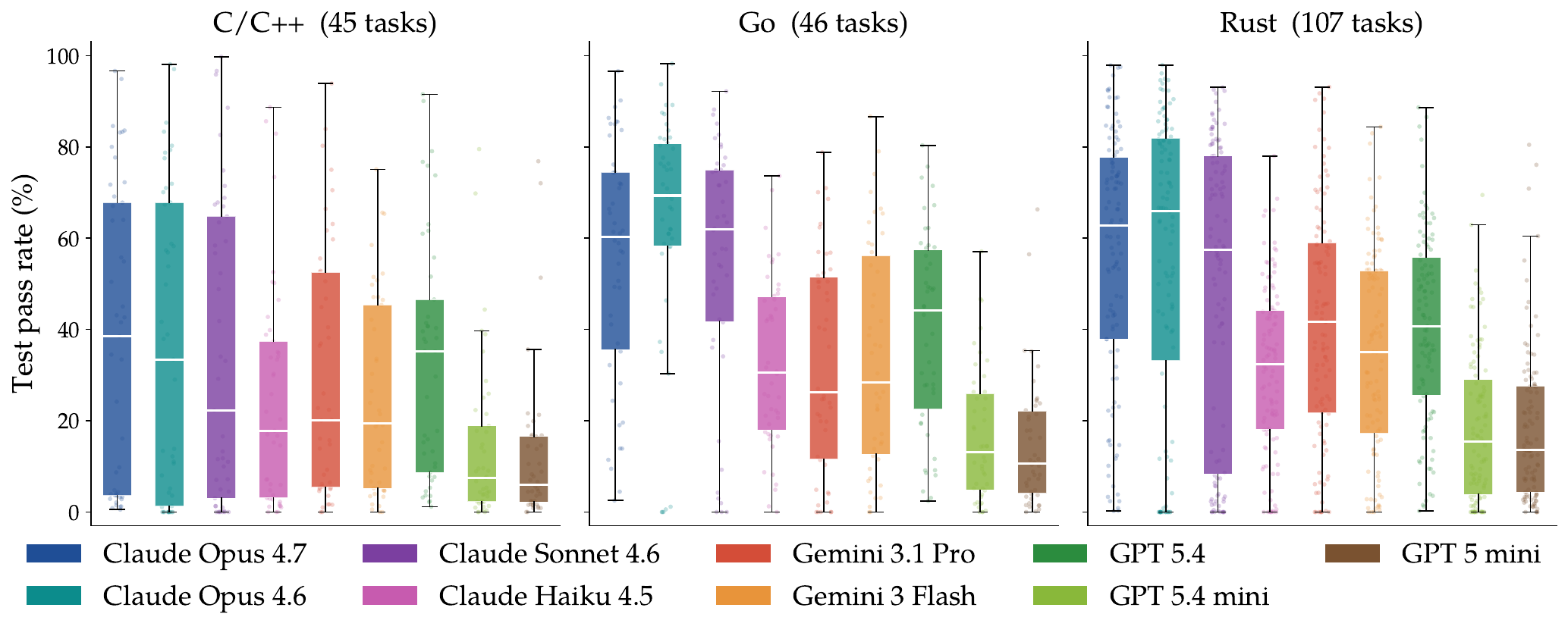}
    \caption{Distribution of per-task test pass rates by reference language across all models. Each box spans the interquartile range with the median shown in white; individual task scores are overlaid as jittered points. Two tasks whose reference language is the sole representative of its family (one Haskell, one Java) are omitted.}
    \label{fig:scores_by_language}
\end{figure}

\textbf{Across all models, performance on C/C++ tasks is substantially lower than Go and Rust.}
We break down pass rates by reference language to understand whether task difficulty varies across language families.
Figure~\ref{fig:scores_by_language} shows per-model box plots for each of the three reference languages.
Rust and Go tasks yield similar average pass rates (38.5\% and 38.4\% respectively), while C/C++ tasks lag behind at 27.7\%.
The model ranking is largely preserved across languages, but C/C++ compresses the distribution downward: even the strongest models see median pass rates drop by 15--20 percentage points relative to Go.
The gap likely reflects the greater complexity of the C/C++ codebases in \bench{} (which include projects such as FFmpeg, GCC, and DuckDB) rather than an inherent language disadvantage.

\begin{figure}[ht]
    \centering
    \includegraphics[width=\textwidth]{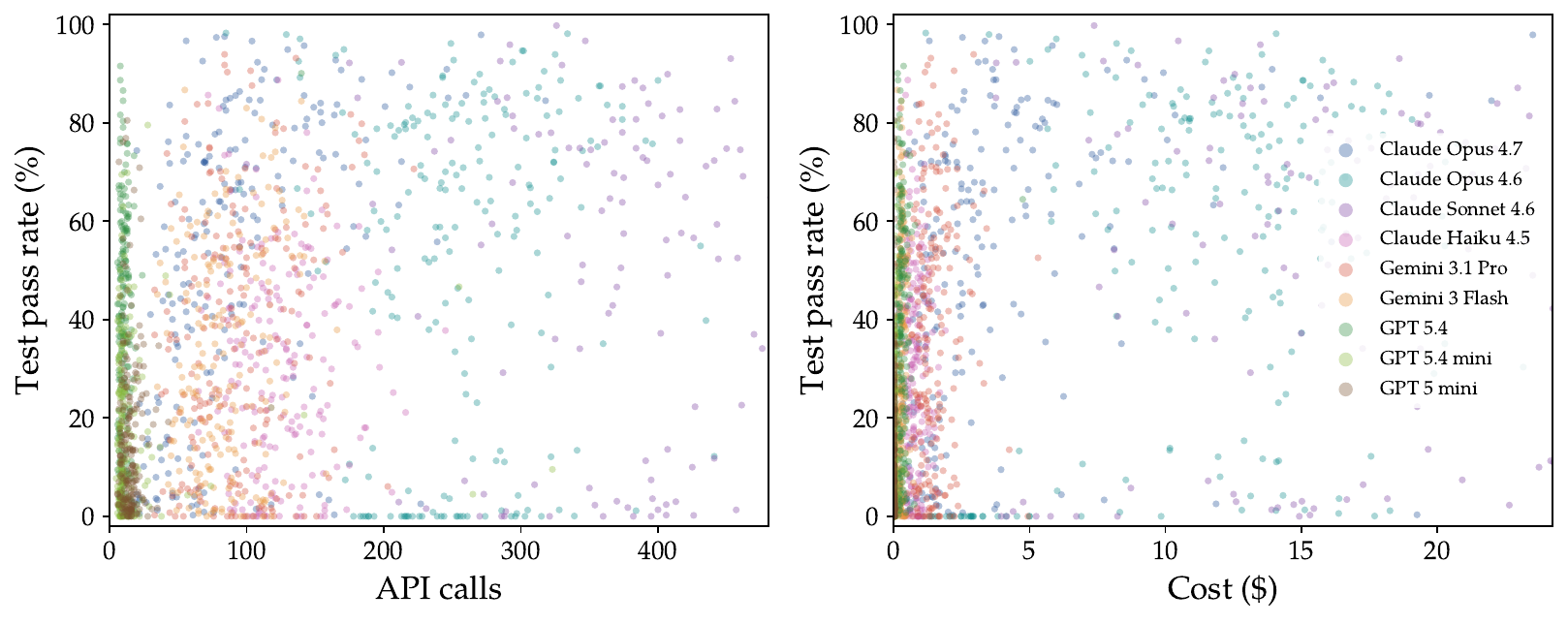}
    \caption{Per-instance test pass rate vs.\ API calls (left) and cost (right). Each point is one (task, model) run. The x-axes are clipped at the 95th percentile for readability.}
    \label{fig:cost_turn_correlation}
\end{figure}

\textbf{More turns spent does not correlate with improved scores.}
We plot per-instance test pass rate against API calls and cost to understand whether additional compute translates to better outcomes.
Figure~\ref{fig:cost_turn_correlation} shows scatter plots across all 1,788 runs, colored by model.
The per-instance Pearson correlations are weak ($r{=}0.27$ for API calls, $r{=}0.21$ for cost), and the scatter plots show no clear trend: high-scoring runs appear at all turn counts and cost levels, while many expensive runs still score near zero.
The modest positive correlations are largely driven by model-level differences, as stronger models tend to both use more turns and score higher, rather than by within-model scaling.

\begin{figure}[ht]
    \begin{minipage}[t]{0.48\textwidth}
        \centering
        \includegraphics[width=\textwidth]{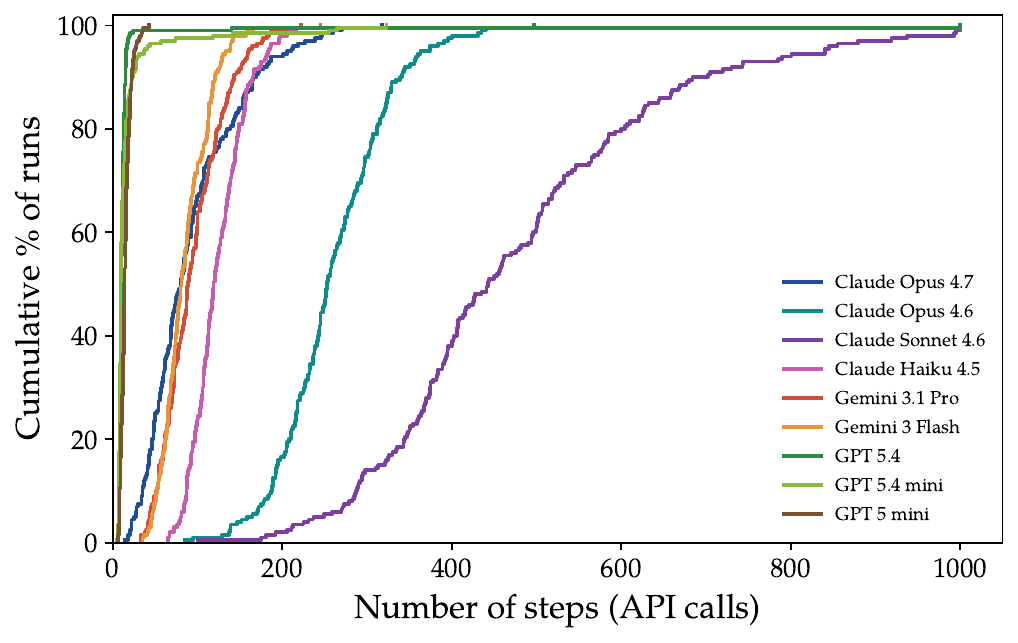}
        \captionof{figure}{Cumulative distribution of trajectory lengths (API calls) per model. Lines that reach 100\% before the 1,000-step limit indicate that all runs for that model submitted voluntarily.}
        \label{fig:turn_count_cdf}
    \end{minipage}
    \hfill
    \begin{minipage}[t]{0.48\textwidth}
        \centering
        \includegraphics[width=\textwidth]{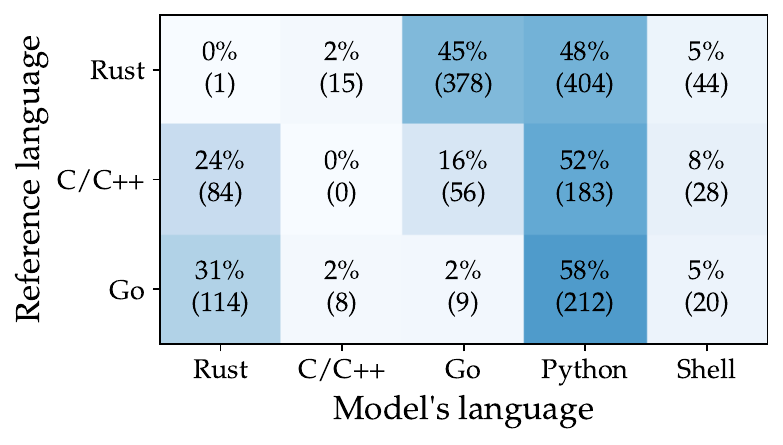}
        \captionof{figure}{Language confusion matrix under the different-language constraint.
        Compared to the default setting (Figure~\ref{fig:language_match}), the diagonal is
        nearly empty and Python becomes the dominant choice across all reference languages.}
        \label{fig:language_match_diff_lang}
    \end{minipage}
\end{figure}

\textbf{Models vary by over an order of magnitude in how many steps they use, falling into three distinct tiers.}
We examine the cumulative distribution of trajectory lengths to understand how models differ in compute allocation.
Figure~\ref{fig:turn_count_cdf} shows CDFs of API calls per run for each model.
The GPT family is the most concise: GPT~5.4, GPT~5.4~mini, and GPT~5~mini all finish 90\% of tasks within 25 steps, with medians of 10, 9, and 14 respectively.
Opus~4.7, Gemini~3.1 Pro, Gemini~3 Flash, and Haiku~4.5 form a middle tier with medians between 80 and 119 steps.
Opus~4.6 and Sonnet~4.6 are clear outliers: Opus~4.6 has a median of 253 steps and Sonnet~4.6 reaches 443, with its 95th percentile at 840 steps approaching the 1,000-step cap.
Combined with the weak score-vs-turns correlation (Figure~\ref{fig:cost_turn_correlation}), these differences appear to reflect model-specific interaction styles rather than a strategy that pays off in higher scores.

\textbf{Models comply with the different-language constraint, defaulting overwhelmingly to Python.}
We examine the language confusion matrix under the different-language ablation setting to verify that models respect the constraint and to see which alternative languages they prefer.
Figure~\ref{fig:language_match_diff_lang} shows the results.
The near-empty diagonal confirms that models largely comply, abandoning same-language reimplementations in favor of alternatives.
Python dominates as the fallback across all three reference languages, accounting for 48--58\% of runs per row and 51\% overall, compared to 36\% in the default setting (Figure~\ref{fig:language_match}).
Despite this heavy skew, the different-language runs achieve only modestly lower pass rates overall (\S\ref{sec:results}), providing empirical evidence that cross-language reimplementation is practical and reinforcing the theoretical argument laid out in \S\ref{appx:benchmark:inference:impossible}.

\begin{figure}[ht]
    \centering
    \includegraphics[width=\textwidth]{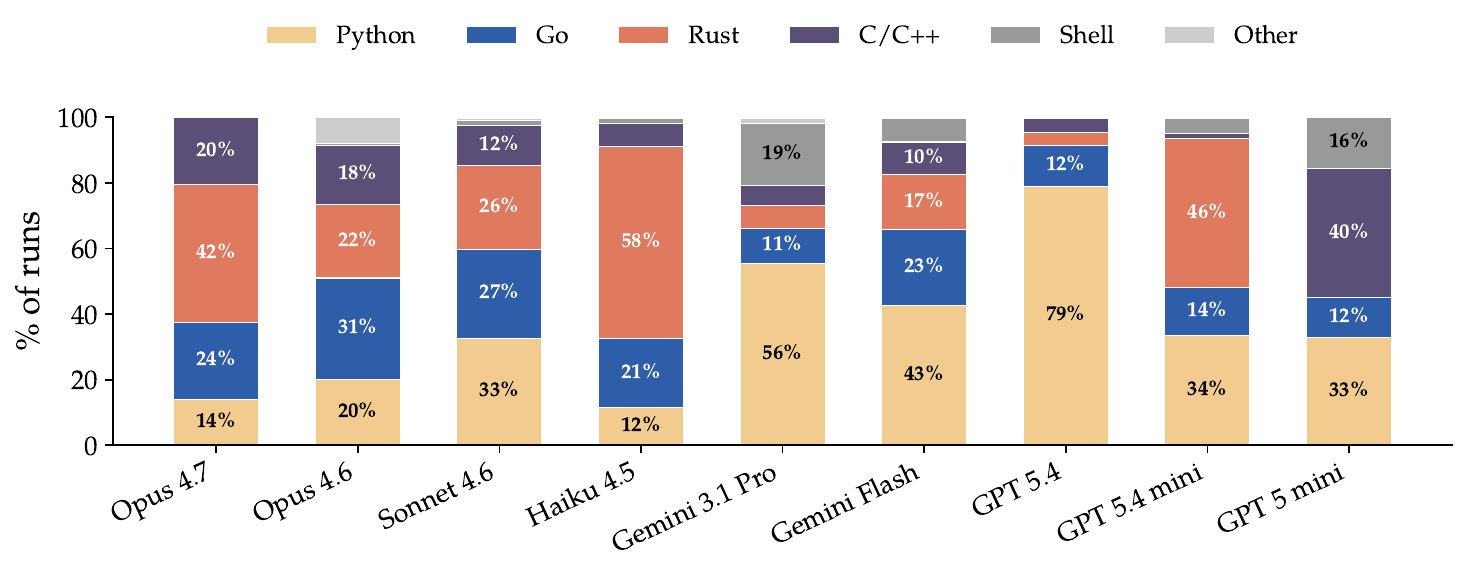}
    \caption{Implementation language chosen by each model, as a percentage of runs.
    Models are free to choose any language; preferences vary significantly across models.}
    \label{fig:language_preference}
\end{figure}

\textbf{Models exhibit distinct language preferences, ranging from Python-dominant to broadly distributed.}
Since models are free to choose any language for their reimplementation, we examine which languages each model gravitates toward.
Figure~\ref{fig:language_preference} shows the breakdown across all runs.
GPT~5.4 is the most skewed, writing 79\% of its solutions in Python, while Gemini~3.1 Pro and Gemini~3 Flash also lean heavily toward Python (56\% and 43\%).
Claude Opus~4.7 and Opus~4.6 instead favor Rust and Go, with Python accounting for only 14\% and 20\% of their runs respectively.
Sonnet~4.6 produces the most balanced distribution, with meaningful fractions across Rust, Go, Python, and C/C++.
These preferences likely reflect differences in training data composition and instruction tuning rather than task-level signals, as the same tasks elicit different language choices from different models.

\begin{figure}[ht]
    \begin{minipage}[t]{0.48\textwidth}
        \centering
        \includegraphics[width=\textwidth]{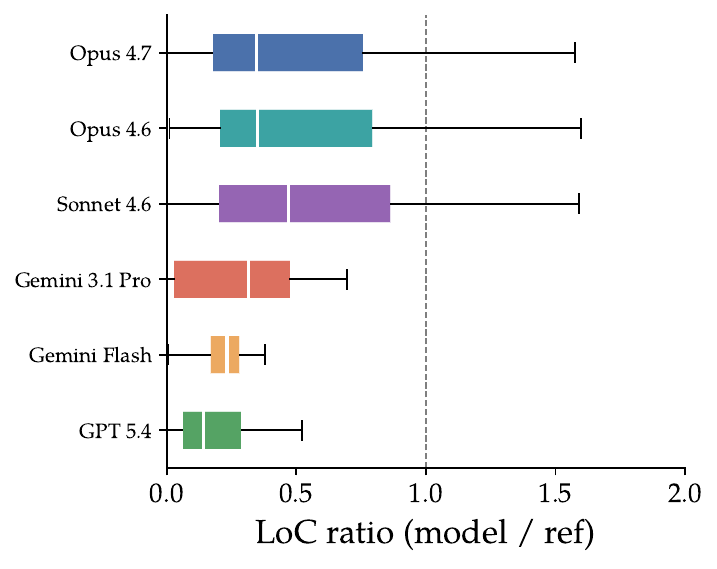}
        \captionof{figure}{LoC ratio (model / reference) per model for solutions passing $\geq$75\% of tests. Dashed line marks parity.}
        \label{fig:loc_ratio_per_model}
    \end{minipage}
    \hfill
    \begin{minipage}[t]{0.48\textwidth}
        \centering
        \includegraphics[width=\textwidth]{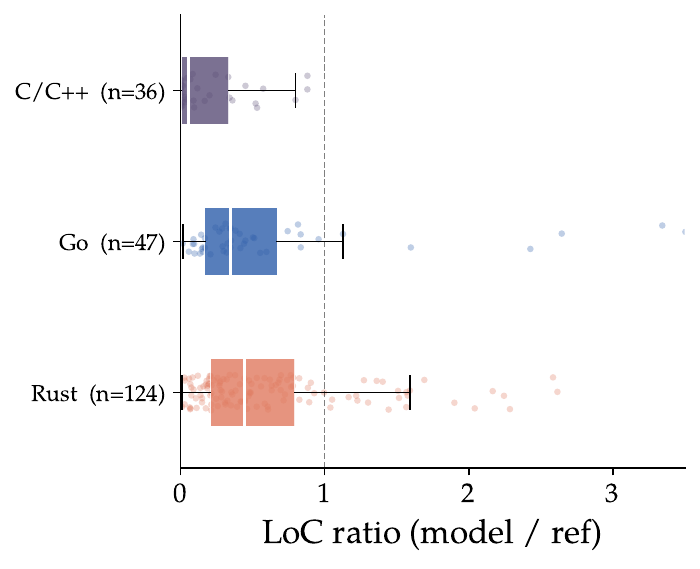}
        \captionof{figure}{LoC ratio by reference language for solutions passing $\geq$75\% of tests. C/C++ references see the largest gap, while Rust references are closest to parity.}
        \label{fig:loc_ratio_by_lang}
    \end{minipage}
\end{figure}

\textbf{Model solutions are consistently shorter than the reference, but the gap varies across models.}
We compare the lines of code in model solutions to the reference codebase for runs passing at least 75\% of tests, to understand how much code models need to reproduce equivalent functionality.
Figure~\ref{fig:loc_ratio_per_model} shows the LoC ratio per model.
All models fall well below parity (the dashed line at 1.0), with medians ranging from roughly 0.15 for Gemini Flash and GPT~5.4 to 0.35 for Opus~4.7 and Sonnet~4.6.
The wide whiskers for Opus~4.7, Opus~4.6, and Sonnet~4.6 indicate that these models occasionally produce solutions approaching or exceeding the reference in length, while Gemini Flash and GPT~5.4 are tightly concentrated at the low end.

\textbf{The LoC ratio gap is largest for C/C++ references and smallest for Rust.}
We further break down LoC ratios by reference language to disentangle language-level effects.
Figure~\ref{fig:loc_ratio_by_lang} shows the results.
C/C++ references exhibit the lowest ratios, with a median around 0.2, consistent with models frequently rewriting these tasks in higher-level languages like Python.
Rust references are closest to parity, with a median near 0.5 and several outliers exceeding 1.0, likely because models that stay in Rust retain similar verbosity to the original.
Go falls in between, with a median around 0.4.

\begin{figure}[ht]
    \centering
    \includegraphics[width=\textwidth]{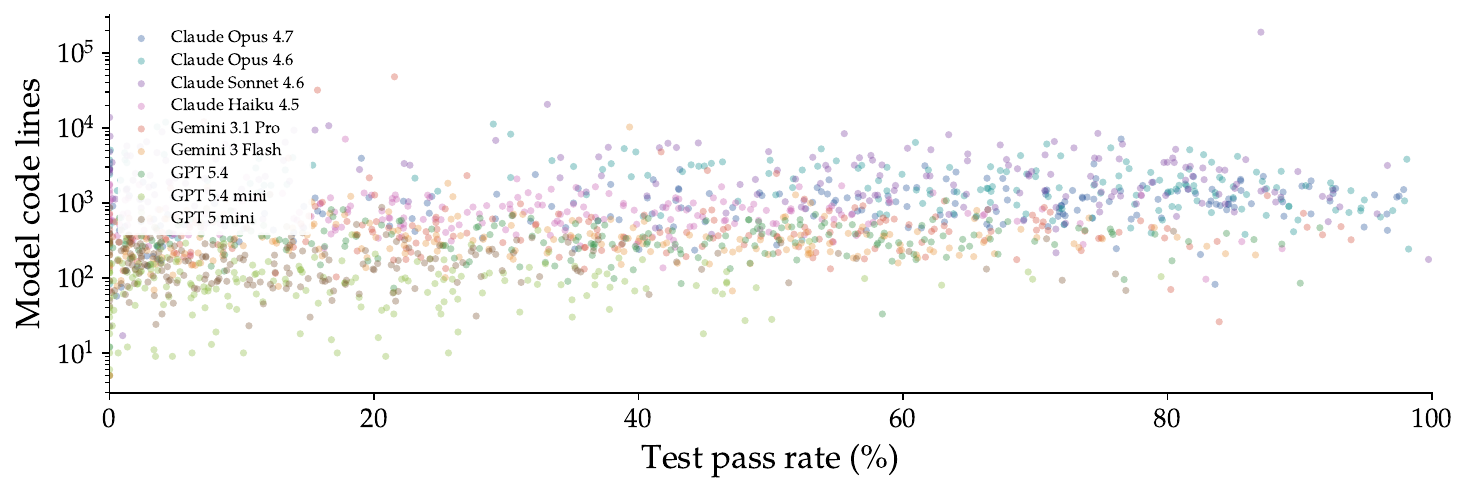}
    \caption{Test pass rate vs.\ model code lines across all 1,800 runs. Points are colored by model.}
    \label{fig:score_vs_loc}
\end{figure}

\textbf{Higher-scoring solutions tend to contain more code, but code volume alone does not guarantee high scores.}
We plot test pass rate against model code lines to examine whether there is a relationship between solution size and performance.
Figure~\ref{fig:score_vs_loc} shows the result across all 1,788 runs on a log-scaled y-axis.
At the low end of the score spectrum, solutions span a wide range of sizes, from under 10 lines to over 10,000.
As pass rates increase, the floor rises: solutions scoring above 75\% cluster between roughly 200 and 10,000 lines, suggesting that a minimum level of implementation completeness is necessary to pass most tests.
However, many large solutions still score poorly, and the overall relationship is noisy.
\FloatBarrier

\newpage
\newpage
\section{Miscellaneous}
\label{appx:miscellaneous}

\subsection{Repository Index}
\label{appx:miscellaneous:repositories}

Table~\ref{tab:repository_list} provides a complete listing of all 200 repositories included in \bench{}, along with a brief description of each project's functionality, its difficulty score, and difficulty label (Easy, Medium, or Hard) as defined in \S\ref{appx:benchmark:statistics}.
As a reminder, the thresholds are: Easy ($<$2), Medium (2 $\leq \text{score}<$ 4), and Hard ($\geq$4).
Task difficulty scores fall in the range of 0 to 10.

{\small
\begin{longtable}{p{0.17\textwidth}p{0.53\textwidth}rc}
\toprule
\textbf{Repository} & \textbf{Description} & \textbf{Score} & \textbf{Difficulty} \\
\midrule
\endfirsthead
\toprule
\textbf{Repository} & \textbf{Description} & \textbf{Score} & \textbf{Difficulty} \\
\midrule
\endhead
\midrule
\multicolumn{4}{r}{\textit{Continued on next page}} \\
\bottomrule
\endfoot
\bottomrule
\\[-0.5em]
\caption{Complete list of repositories in \bench{} with brief descriptions and difficulty scores. $^\dagger$\,Networking tool; tested over localhost (see \S\ref{appx:benchmark:inference:impossible}).}
\label{tab:repository_list}
\endlastfoot
\texttt{abishekvashok/}\newline\hspace{1em}\texttt{cmatrix} & Terminal screensaver that simulates the falling text from The Matrix & 1.8 & \colorbox{green!15}{\strut Easy} \\
\texttt{agourlay/}\newline\hspace{1em}\texttt{zip-password-finder} & Brute-force password recovery tool for protected ZIP archives & 2.3 & \colorbox{yellow!25}{\strut Medium} \\
\texttt{ajeetdsouza/zoxide} & A smarter cd command. Supports all major shells. & 2.6 & \colorbox{yellow!25}{\strut Medium} \\
\texttt{alecthomas/chroma} & A general purpose syntax highlighter in pure Go  & 2.8 & \colorbox{yellow!25}{\strut Medium} \\
\texttt{alexpovel/srgn} & Syntax-aware code search and replacement tool supporting multiple languages & 3.7 & \colorbox{yellow!25}{\strut Medium} \\
\texttt{altdesktop/}\newline\hspace{1em}\texttt{i3-style}$^\dagger$ & Applies color themes to i3 window manager configuration files & 2.1 & \colorbox{yellow!25}{\strut Medium} \\
\texttt{ammarabouzor/}\newline\hspace{1em}\texttt{tui-journal} & Terminal-based journaling application with a text user interface & 3.4 & \colorbox{yellow!25}{\strut Medium} \\
\texttt{anordal/}\newline\hspace{1em}\texttt{shellharden} & Bash syntax analyzer that suggests quoting and safety corrections & 1.6 & \colorbox{green!15}{\strut Easy} \\
\texttt{antonmedv/fx} & Terminal-based JSON viewer and interactive processor & 3.4 & \colorbox{yellow!25}{\strut Medium} \\
\texttt{antonmedv/walk} & Terminal file manager with interactive directory navigation & 2.3 & \colorbox{yellow!25}{\strut Medium} \\
\texttt{ariga/atlas} & Declarative database schema migration tool using schema-as-code workflows & 4.2 & \colorbox{red!15}{\strut Hard} \\
\texttt{arq5x/bedtools2} & bedtools - the swiss army knife for genome arithmetic & 3.0 & \colorbox{yellow!25}{\strut Medium} \\
\texttt{arthursonzogni/}\newline\hspace{1em}\texttt{json-tui} & Terminal user interface for browsing and navigating JSON data & 1.8 & \colorbox{green!15}{\strut Easy} \\
\texttt{ast-grep/ast-grep} & A CLI tool for code structural search, lint and rewriting. Written in Rust & 4.5 & \colorbox{red!15}{\strut Hard} \\
\texttt{astaxie/bat} & Go-based HTTP client for the command line, similar to cURL & 1.3 & \colorbox{green!15}{\strut Easy} \\
\texttt{astro/deadnix} & Static analyzer that detects unused code in Nix expressions & 2.2 & \colorbox{yellow!25}{\strut Medium} \\
\texttt{axodotdev/oranda} & Static site generator for creating landing pages for developer tools & 3.7 & \colorbox{yellow!25}{\strut Medium} \\
\texttt{bellard/quickjs} & Public repository of the QuickJS Javascript Engine. & 3.0 & \colorbox{yellow!25}{\strut Medium} \\
\texttt{bensadeh/tailspin} & Log file viewer with syntax highlighting for common patterns & 3.1 & \colorbox{yellow!25}{\strut Medium} \\
\texttt{blacknon/hwatch}$^\dagger$ & File-watching alternative to the watch command that records and diffs output history & 3.4 & \colorbox{yellow!25}{\strut Medium} \\
\texttt{blake3-team/blake3} & the official Rust and C implementations of the BLAKE3 cryptographic hash function & 3.3 & \colorbox{yellow!25}{\strut Medium} \\
\texttt{bootandy/dust} & Disk usage analyzer that displays directory sizes as a visual tree & 2.9 & \colorbox{yellow!25}{\strut Medium} \\
\texttt{boyter/scc} & Sloc, Cloc and Code: scc is a very fast accurate code counter with complexity calculations and COCOMO estimates written in pure Go & 4.6 & \colorbox{red!15}{\strut Hard} \\
\texttt{brocode/fblog} & Command-line viewer for JSON-formatted log files & 2.3 & \colorbox{yellow!25}{\strut Medium} \\
\texttt{burntsushi/ripgrep} & Fast recursive regex search tool that respects gitignore rules & 3.7 & \colorbox{yellow!25}{\strut Medium} \\
\texttt{burntsushi/xsv} & High-performance command-line toolkit for working with CSV files & 3.0 & \colorbox{yellow!25}{\strut Medium} \\
\texttt{byron/dua-cli} & Disk usage analyzer with interactive mode for reviewing and deleting files & 3.3 & \colorbox{yellow!25}{\strut Medium} \\
\texttt{canop/broot} & Terminal-based directory tree navigator and file manager & 4.3 & \colorbox{red!15}{\strut Hard} \\
\texttt{canop/rhit} & Nginx log file analyzer and statistics viewer & 2.8 & \colorbox{yellow!25}{\strut Medium} \\
\texttt{cheat/cheat} & cheat allows you to create and view interactive cheatsheets on the command-line. It was designed to help remind *nix system administrators of options for commands that they use frequently, but not frequently enough to remember. & 4.5 & \colorbox{red!15}{\strut Hard} \\
\texttt{chirlu/sox} & SoX, Swiss Army knife of sound processing & 2.8 & \colorbox{yellow!25}{\strut Medium} \\
\texttt{chmln/handlr}$^\dagger$ & Command-line tool for managing default applications on Linux via MIME types & 2.3 & \colorbox{yellow!25}{\strut Medium} \\
\texttt{chmln/sd} & Command-line find-and-replace tool designed as a simpler alternative to sed & 2.1 & \colorbox{yellow!25}{\strut Medium} \\
\texttt{clog-tool/clog-cli} & Changelog generator that parses conventional Git commit messages & 1.5 & \colorbox{green!15}{\strut Easy} \\
\texttt{cmatsuoka/figlet} & Generates large ASCII art text banners from input strings & 1.8 & \colorbox{green!15}{\strut Easy} \\
\texttt{codesnap-rs/}\newline\hspace{1em}\texttt{codesnap} & Generates stylized code snippet images from source files & 3.3 & \colorbox{yellow!25}{\strut Medium} \\
\texttt{cordx56/rustowl} & Visualizes ownership and lifetime annotations in Rust source code & 3.3 & \colorbox{yellow!25}{\strut Medium} \\
\texttt{crowdagger/}\newline\hspace{1em}\texttt{crowbook} & Converts Markdown books to HTML, LaTeX, PDF, and EPUB formats & 3.4 & \colorbox{yellow!25}{\strut Medium} \\
\texttt{cslarsen/jp2a} & Converts JPEG images to ASCII art for terminal display & 1.7 & \colorbox{green!15}{\strut Easy} \\
\texttt{cweill/gotests} & Generates Go test function boilerplate from source code signatures & 2.5 & \colorbox{yellow!25}{\strut Medium} \\
\texttt{dalance/amber} & Command-line code search and replacement tool with regex support & 2.9 & \colorbox{yellow!25}{\strut Medium} \\
\texttt{dandavison/delta} & Syntax-highlighting pager for git diff, grep, and blame output & 4.0 & \colorbox{red!15}{\strut Hard} \\
\texttt{danmar/cppcheck} & Static analysis tool for detecting bugs in C and C++ code & 3.9 & \colorbox{yellow!25}{\strut Medium} \\
\texttt{direnv/direnv} & Automatically loads and unloads environment variables per directory & 2.5 & \colorbox{yellow!25}{\strut Medium} \\
\texttt{doxygen/doxygen} & Documentation generator for C++, C, Java, and other languages & 5.1 & \colorbox{red!15}{\strut Hard} \\
\texttt{drew-alleman/}\newline\hspace{1em}\texttt{datasurgeon} & Extracts structured data such as IPs, emails, and hashes from text & 2.0 & \colorbox{green!15}{\strut Easy} \\
\texttt{ducaale/xh}$^\dagger$ & Command-line HTTP client with a user-friendly interface & 3.8 & \colorbox{yellow!25}{\strut Medium} \\
\texttt{duckdb/duckdb} & DuckDB is an analytical in-process SQL database management system & 4.7 & \colorbox{red!15}{\strut Hard} \\
\texttt{dundee/gdu} & Fast interactive disk usage analyzer with a terminal interface & 3.5 & \colorbox{yellow!25}{\strut Medium} \\
\texttt{ecumene/rust-sloth} & Software 3D rasterizer that renders graphics in the terminal & 2.5 & \colorbox{yellow!25}{\strut Medium} \\
\texttt{ekzhang/bore}$^\dagger$ & CLI tool for creating network tunnels to expose localhost to the internet & 2.1 & \colorbox{yellow!25}{\strut Medium} \\
\texttt{eliukblau/pixterm} & Renders images in the terminal using ANSI true color escape sequences & 1.7 & \colorbox{green!15}{\strut Easy} \\
\texttt{elkowar/pipr} & Interactive tool for incrementally building Unix shell pipelines & 2.6 & \colorbox{yellow!25}{\strut Medium} \\
\texttt{epistates/treemd} & Terminal Markdown viewer with tree-based structural navigation & 3.9 & \colorbox{yellow!25}{\strut Medium} \\
\texttt{eradman/entr} & Runs specified commands automatically when watched files change & 1.3 & \colorbox{green!15}{\strut Easy} \\
\texttt{esubaalew/run} & Universal multi-language script runner and REPL & 3.3 & \colorbox{yellow!25}{\strut Medium} \\
\texttt{eudoxia0/hashcards} & Plain-text spaced repetition flashcard system for the command line & 3.1 & \colorbox{yellow!25}{\strut Medium} \\
\texttt{facebook/zstd} & Fast lossless compression algorithm and library by Facebook & 3.2 & \colorbox{yellow!25}{\strut Medium} \\
\texttt{facebookresearch/}\newline\hspace{1em}\texttt{fasttext} & Library for fast text representation and classification. & 2.2 & \colorbox{yellow!25}{\strut Medium} \\
\texttt{ffmpeg/ffmpeg} & Multimedia framework for encoding, decoding, transcoding, and streaming audio and video & 4.2 & \colorbox{red!15}{\strut Hard} \\
\texttt{filosottile/age} & A simple, modern and secure encryption tool (and Go library) with small explicit keys, no config options, and UNIX-style composability. & 3.0 & \colorbox{yellow!25}{\strut Medium} \\
\texttt{foriequal0/}\newline\hspace{1em}\texttt{git-trim} & Automatically deletes local Git branches whose remote refs are merged or deleted & 2.9 & \colorbox{yellow!25}{\strut Medium} \\
\texttt{gabotechs/dep-tree} & Visualizes source code file dependencies as a 3D force-directed graph & 3.4 & \colorbox{yellow!25}{\strut Medium} \\
\texttt{ggreer/}\newline\hspace{1em}\texttt{the\_silver\_searcher} & Fast code search tool similar to ack, optimized for large codebases & 2.6 & \colorbox{yellow!25}{\strut Medium} \\
\texttt{git-bahn/git-graph} & Displays Git commit history as a formatted branching graph & 2.9 & \colorbox{yellow!25}{\strut Medium} \\
\texttt{go-critic/}\newline\hspace{1em}\texttt{go-critic} & Opinionated Go source code linter for style and correctness auditing & 3.6 & \colorbox{yellow!25}{\strut Medium} \\
\texttt{google/brotli} & Brotli compression format & 2.9 & \colorbox{yellow!25}{\strut Medium} \\
\texttt{gromacs/gromacs} & Public/backup repository of the GROMACS molecular simulation toolkit.  & 5.6 & \colorbox{red!15}{\strut Hard} \\
\texttt{guumaster/hostctl}$^\dagger$ & Command-line tool for managing /etc/hosts file entries by profile & 2.7 & \colorbox{yellow!25}{\strut Medium} \\
\texttt{hairyhenderson/}\newline\hspace{1em}\texttt{gomplate} & Command-line template rendering tool supporting multiple data sources & 4.0 & \colorbox{red!15}{\strut Hard} \\
\texttt{halitechallenge/}\newline\hspace{1em}\texttt{halite} & AI programming competition framework for building game-playing bots & 2.9 & \colorbox{yellow!25}{\strut Medium} \\
\texttt{hatoo/oha}$^\dagger$ & HTTP load testing tool with real-time terminal animation & 3.6 & \colorbox{yellow!25}{\strut Medium} \\
\texttt{hooklift/gowsdl} & Generates Go client code from WSDL service definitions & 2.0 & \colorbox{yellow!25}{\strut Medium} \\
\texttt{hpjansson/chafa} & Renders images and animations as character art in the terminal & 3.9 & \colorbox{yellow!25}{\strut Medium} \\
\texttt{htop-dev/htop}$^\dagger$ & Interactive terminal-based process viewer and system monitor & 4.0 & \colorbox{yellow!25}{\strut Medium} \\
\texttt{hush-shell/hush} & Unix shell built on the Lua programming language & 3.5 & \colorbox{yellow!25}{\strut Medium} \\
\texttt{incu6us/}\newline\hspace{1em}\texttt{goimports-reviser} & Go import sorting and code formatting tool & 2.5 & \colorbox{yellow!25}{\strut Medium} \\
\texttt{ip7z/7zip} & 7-Zip & 3.5 & \colorbox{yellow!25}{\strut Medium} \\
\texttt{ismaelgv/rnr} & Command-line tool for batch renaming files using regex patterns & 2.5 & \colorbox{yellow!25}{\strut Medium} \\
\texttt{isona/dirble} & Fast web directory and file enumeration scanner & 3.0 & \colorbox{yellow!25}{\strut Medium} \\
\texttt{ivanceras/svgbob} & Convert your ascii diagram scribbles into happy little SVG & 3.2 & \colorbox{yellow!25}{\strut Medium} \\
\texttt{jarun/nnn} & Lightweight and fast terminal file manager & 2.2 & \colorbox{yellow!25}{\strut Medium} \\
\texttt{jesseduffield/}\newline\hspace{1em}\texttt{lazygit} & Terminal user interface for common Git operations & 5.3 & \colorbox{red!15}{\strut Hard} \\
\texttt{jgm/pandoc} & Universal markup converter & 3.1 & \colorbox{yellow!25}{\strut Medium} \\
\texttt{jhspetersson/}\newline\hspace{1em}\texttt{fselect} & Find files with SQL-like queries & 3.9 & \colorbox{yellow!25}{\strut Medium} \\
\texttt{johanneskaufmann/}\newline\hspace{1em}\texttt{html-to-markdown} & Converts HTML content to Markdown with configurable rules & 3.1 & \colorbox{yellow!25}{\strut Medium} \\
\texttt{johnkerl/miller} & Command-line tool for processing structured data in CSV, TSV, and JSON formats & 5.3 & \colorbox{red!15}{\strut Hard} \\
\texttt{jonas/tig} & Text-mode interface for browsing Git repositories & 3.1 & \colorbox{yellow!25}{\strut Medium} \\
\texttt{jqlang/jq} & Command-line processor for querying and transforming JSON data & 3.5 & \colorbox{yellow!25}{\strut Medium} \\
\texttt{jrnxf/thokr} & Terminal typing speed test with result visualization and history logging & 2.1 & \colorbox{yellow!25}{\strut Medium} \\
\texttt{junegunn/fzf} & General-purpose command-line fuzzy finder for interactive filtering & 3.6 & \colorbox{yellow!25}{\strut Medium} \\
\texttt{kaushiksrini/}\newline\hspace{1em}\texttt{parqeye} & Terminal tool for inspecting and previewing Parquet file contents & 2.7 & \colorbox{yellow!25}{\strut Medium} \\
\texttt{kisielk/errcheck} & Go linter that detects unchecked error return values & 1.8 & \colorbox{green!15}{\strut Easy} \\
\texttt{konradsz/igrep} & Interactive grep tool with a terminal user interface & 2.6 & \colorbox{yellow!25}{\strut Medium} \\
\texttt{ksxgithub/}\newline\hspace{1em}\texttt{parallel-disk-usage} & Parallelized directory tree size analyzer for fast disk usage reporting & 3.4 & \colorbox{yellow!25}{\strut Medium} \\
\texttt{kyoh86/richgo} & Enriches Go test output with color and formatting decorations & 2.3 & \colorbox{yellow!25}{\strut Medium} \\
\texttt{kyoheiu/felix} & Terminal file manager with Vim-style key bindings & 3.2 & \colorbox{yellow!25}{\strut Medium} \\
\texttt{lfos/calcurse} & Text-based calendar and scheduling application for the terminal & 2.5 & \colorbox{yellow!25}{\strut Medium} \\
\texttt{lh3/seqtk} & Toolkit for processing sequences in FASTA/Q formats & 1.5 & \colorbox{green!15}{\strut Easy} \\
\texttt{lua/lua} & Reference implementation of the Lua programming language interpreter & 2.7 & \colorbox{yellow!25}{\strut Medium} \\
\texttt{luajit/luajit} & High-performance just-in-time compiler for the Lua programming language & 3.0 & \colorbox{yellow!25}{\strut Medium} \\
\texttt{lymphatus/}\newline\hspace{1em}\texttt{caesium-clt} & Command-line image compression tool supporting lossy and lossless modes & 2.4 & \colorbox{yellow!25}{\strut Medium} \\
\texttt{lz4/lz4} & Extremely fast lossless compression algorithm and library & 2.5 & \colorbox{yellow!25}{\strut Medium} \\
\texttt{madler/pigz} & Parallel implementation of gzip for multi-core processors & 2.0 & \colorbox{yellow!25}{\strut Medium} \\
\texttt{mfridman/tparse} & Summarizes and formats Go test output for terminals and CI pipelines & 2.3 & \colorbox{yellow!25}{\strut Medium} \\
\texttt{mgdm/htmlq} & Command-line tool for extracting content from HTML using CSS selectors & 1.5 & \colorbox{green!15}{\strut Easy} \\
\texttt{mgechev/revive} & Fast and configurable Go linter as a drop-in replacement for golint & 3.6 & \colorbox{yellow!25}{\strut Medium} \\
\texttt{mibk/dupl} & Detects duplicate code fragments in Go source files & 1.3 & \colorbox{green!15}{\strut Easy} \\
\texttt{mikefarah/yq} & yq is a portable command-line YAML, JSON, XML, CSV, TOML, HCL  and properties processor & 4.0 & \colorbox{red!15}{\strut Hard} \\
\texttt{miserlou/loop}$^\dagger$ & Command-line utility for repeating commands with intervals and counters & 1.6 & \colorbox{green!15}{\strut Easy} \\
\texttt{mkj/dropbear}$^\dagger$ & Lightweight SSH server and client implementation & 4.0 & \colorbox{yellow!25}{\strut Medium} \\
\texttt{mookid/diffr} & Side-by-side diff viewer with word-level highlighting & 2.0 & \colorbox{yellow!25}{\strut Medium} \\
\texttt{multiprocessio/dsq} & Runs SQL queries against JSON, CSV, Excel, and Parquet files from the command line & 1.9 & \colorbox{green!15}{\strut Easy} \\
\texttt{nachoparker/dutree} & Disk usage analyzer that displays results as a colored directory tree & 1.7 & \colorbox{green!15}{\strut Easy} \\
\texttt{naggie/dstask} & Git-backed terminal task and note manager with Markdown support & 3.0 & \colorbox{yellow!25}{\strut Medium} \\
\texttt{nikoladucak/}\newline\hspace{1em}\texttt{caps-log} & Terminal-based journaling application with a calendar interface & 3.0 & \colorbox{yellow!25}{\strut Medium} \\
\texttt{nikolassv/bartib} & Command-line time tracker that stores activity logs as plain text & 2.5 & \colorbox{yellow!25}{\strut Medium} \\
\texttt{ninja-build/ninja} & Small and fast build system focused on incremental compilation speed & 3.0 & \colorbox{yellow!25}{\strut Medium} \\
\texttt{noborus/ov} & Feature-rich terminal pager for viewing text files and command output & 3.9 & \colorbox{yellow!25}{\strut Medium} \\
\texttt{noborus/trdsql} & Executes SQL queries on CSV, JSON, YAML, and other tabular file formats & 3.4 & \colorbox{yellow!25}{\strut Medium} \\
\texttt{nukesor/pueue} & Manage your shell commands. & 4.0 & \colorbox{yellow!25}{\strut Medium} \\
\texttt{nuta/nsh} & POSIX-compatible command-line shell with fish-like interactive features & 3.3 & \colorbox{yellow!25}{\strut Medium} \\
\texttt{o2sh/onefetch} & Displays Git repository summary and statistics in the terminal & 3.2 & \colorbox{yellow!25}{\strut Medium} \\
\texttt{ogham/dog}$^\dagger$ & Command-line DNS lookup client with colorized output & 2.9 & \colorbox{yellow!25}{\strut Medium} \\
\texttt{oppiliappan/eva} & Terminal calculator REPL similar to bc with expression evaluation & 2.0 & \colorbox{yellow!25}{\strut Medium} \\
\texttt{oppiliappan/statix} & Linter and diagnostic tool for the Nix programming language & 3.0 & \colorbox{yellow!25}{\strut Medium} \\
\texttt{orf/gping}$^\dagger$ & Ping utility that displays response times as a live terminal graph & 2.4 & \colorbox{yellow!25}{\strut Medium} \\
\texttt{osgeo/gdal} & GDAL is an open source MIT licensed translator library for raster and vector geospatial data formats. & 5.4 & \colorbox{red!15}{\strut Hard} \\
\texttt{osgeo/proj} & PROJ - Cartographic Projections and Coordinate Transformations Library & 4.4 & \colorbox{red!15}{\strut Hard} \\
\texttt{paradigmxyz/solar} & Modular Solidity compiler written in Rust for fast compilation & 5.2 & \colorbox{red!15}{\strut Hard} \\
\texttt{parcel-bundler/}\newline\hspace{1em}\texttt{lightningcss} & High-performance CSS parser, transformer, bundler, and minifier & 4.5 & \colorbox{red!15}{\strut Hard} \\
\texttt{peco/peco} & Interactive line filtering tool for terminal pipelines & 3.2 & \colorbox{yellow!25}{\strut Medium} \\
\texttt{pemistahl/grex} & Generates regular expressions from user-provided example strings & 3.1 & \colorbox{yellow!25}{\strut Medium} \\
\texttt{php/php-src} & Source code of the PHP programming language interpreter & 5.1 & \colorbox{red!15}{\strut Hard} \\
\texttt{pier-cli/pier} & Organizes and runs short reusable shell scripts from a central registry & 2.3 & \colorbox{yellow!25}{\strut Medium} \\
\texttt{pls-rs/pls} & Modern directory listing tool with formatting and metadata display & 3.1 & \colorbox{yellow!25}{\strut Medium} \\
\texttt{psampaz/}\newline\hspace{1em}\texttt{go-mod-outdated} & Reports outdated dependencies in Go module projects & 1.1 & \colorbox{green!15}{\strut Easy} \\
\texttt{quinn-rs/quinn}$^\dagger$ & Async-compatible QUIC protocol implementation in Rust & 4.2 & \colorbox{red!15}{\strut Hard} \\
\texttt{raviqqe/muffet}$^\dagger$ & Fast recursive website link checker for detecting broken URLs & 2.9 & \colorbox{yellow!25}{\strut Medium} \\
\texttt{rbakbashev/elfcat} & Visualizes ELF binary structure by generating annotated HTML output & 1.4 & \colorbox{green!15}{\strut Easy} \\
\texttt{rcoh/angle-grinder} & Slice and dice logs on the command line & 3.4 & \colorbox{yellow!25}{\strut Medium} \\
\texttt{rhysd/kiro-editor} & Small terminal text editor with UTF-8 support written in Rust & 2.7 & \colorbox{yellow!25}{\strut Medium} \\
\texttt{riquito/tuc} & Column-cutting tool with advanced field selection beyond cut & 2.9 & \colorbox{yellow!25}{\strut Medium} \\
\texttt{robertdavidgraham/}\newline\hspace{1em}\texttt{masscan}$^\dagger$ & Asynchronous TCP port scanner capable of scanning the entire internet rapidly & 2.8 & \colorbox{yellow!25}{\strut Medium} \\
\texttt{rochacbruno/}\newline\hspace{1em}\texttt{marmite} & Static site generator that builds blogs from Markdown files & 3.7 & \colorbox{yellow!25}{\strut Medium} \\
\texttt{rs/curlie}$^\dagger$ & Command-line HTTP client combining curl functionality with a simpler interface & 1.5 & \colorbox{green!15}{\strut Easy} \\
\texttt{rs/jplot} & Real-time JSON and expvar data plotting tool for iTerm2 & 2.0 & \colorbox{green!15}{\strut Easy} \\
\texttt{rust-embedded/}\newline\hspace{1em}\texttt{svd2rust} & Generates Rust register map structs from SVD hardware description files & 3.3 & \colorbox{yellow!25}{\strut Medium} \\
\texttt{rust-ethereum/}\newline\hspace{1em}\texttt{ethabi} & Encodes and decodes Ethereum smart contract ABI invocations & 3.2 & \colorbox{yellow!25}{\strut Medium} \\
\texttt{rust-lang/mdbook} & Generates online books from Markdown source files & 3.6 & \colorbox{yellow!25}{\strut Medium} \\
\texttt{rvben/rumdl} & Markdown linter and formatter written in Rust & 5.1 & \colorbox{red!15}{\strut Hard} \\
\texttt{samtools/samtools} & Tools (written in C using htslib) for manipulating next-generation sequencing data & 3.1 & \colorbox{yellow!25}{\strut Medium} \\
\texttt{sayanarijit/xplr} & Extensible terminal file explorer with a scriptable plugin system & 3.6 & \colorbox{yellow!25}{\strut Medium} \\
\texttt{sclevine/yj} & Converts between YAML, TOML, JSON, and HCL configuration formats & 1.8 & \colorbox{green!15}{\strut Easy} \\
\texttt{segmentio/chamber} & CLI tool for managing application secrets via AWS SSM Parameter Store & 3.1 & \colorbox{yellow!25}{\strut Medium} \\
\texttt{sharkdp/bat} & A cat(1) clone with wings. & 5.6 & \colorbox{red!15}{\strut Hard} \\
\texttt{sharkdp/fd} & Fast and user-friendly alternative to the find command & 3.2 & \colorbox{yellow!25}{\strut Medium} \\
\texttt{sharkdp/hexyl} & Command-line hex viewer with colored output & 2.5 & \colorbox{yellow!25}{\strut Medium} \\
\texttt{sharkdp/hyperfine} & A command-line benchmarking tool & 3.0 & \colorbox{yellow!25}{\strut Medium} \\
\texttt{sharkdp/pastel} & Command-line tool for generating, converting, and manipulating colors & 2.8 & \colorbox{yellow!25}{\strut Medium} \\
\texttt{shashwatah/jot} & Minimal command-line note-taking tool for quick capture & 2.0 & \colorbox{yellow!25}{\strut Medium} \\
\texttt{sheepla/pingu} & Ping wrapper that displays results with a Pingu-themed animation & 0.9 & \colorbox{green!15}{\strut Easy} \\
\texttt{sibprogrammer/xq} & Command-line XML and HTML formatter and content extractor & 2.2 & \colorbox{yellow!25}{\strut Medium} \\
\texttt{sigoden/argc} & A Bash CLI framework, also a Bash command runner. & 3.4 & \colorbox{yellow!25}{\strut Medium} \\
\texttt{simeg/eureka} & CLI tool for quickly capturing and storing ideas from the terminal & 2.3 & \colorbox{yellow!25}{\strut Medium} \\
\texttt{sirwart/ripsecrets} & Pre-commit scanner that prevents secret keys from entering source code & 2.0 & \colorbox{green!15}{\strut Easy} \\
\texttt{sitkevij/hex} & Command-line hex dump viewer written in Rust & 2.0 & \colorbox{yellow!25}{\strut Medium} \\
\texttt{skeema/skeema} & Declarative schema management tool for MySQL and MariaDB using pure SQL & 4.3 & \colorbox{red!15}{\strut Hard} \\
\texttt{sqlite/sqlite} & Official Git mirror of the SQLite source tree & 3.7 & \colorbox{yellow!25}{\strut Medium} \\
\texttt{sstadick/hck} & Fast column-extraction tool as an alternative to cut & 2.8 & \colorbox{yellow!25}{\strut Medium} \\
\texttt{stacked-git/stgit} & Manages a stack of patches on top of Git branches & 4.0 & \colorbox{red!15}{\strut Hard} \\
\texttt{stathissideris/}\newline\hspace{1em}\texttt{ditaa} & ditaa is a small command-line utility that can convert diagrams drawn using ascii art ('drawings' that contain characters that resemble lines like | / - ), into proper bitmap graphics. & 2.1 & \colorbox{yellow!25}{\strut Medium} \\
\texttt{stranger6667/}\newline\hspace{1em}\texttt{jsonschema} & High-performance JSON Schema validation library for Rust & 4.3 & \colorbox{red!15}{\strut Hard} \\
\texttt{svenstaro/genact} & A nonsense activity generator & 2.9 & \colorbox{yellow!25}{\strut Medium} \\
\texttt{svenstaro/}\newline\hspace{1em}\texttt{miniserve}$^\dagger$ & Simple command-line HTTP file server for quick local file sharing & 3.5 & \colorbox{yellow!25}{\strut Medium} \\
\texttt{tarka/xcp} & Extended file copy tool with progress display and parallel operations & 2.9 & \colorbox{yellow!25}{\strut Medium} \\
\texttt{thezoraiz/}\newline\hspace{1em}\texttt{ascii-image-converter} & Converts images to ASCII and Braille art for terminal display & 2.4 & \colorbox{yellow!25}{\strut Medium} \\
\texttt{tinycc/tinycc} & Unofficial mirror of mob development branch & 3.1 & \colorbox{yellow!25}{\strut Medium} \\
\texttt{tomarrell/}\newline\hspace{1em}\texttt{wrapcheck} & Go linter that checks whether errors from external packages are wrapped & 2.5 & \colorbox{yellow!25}{\strut Medium} \\
\texttt{tomnomnom/gron} & Make JSON greppable! & 2.2 & \colorbox{yellow!25}{\strut Medium} \\
\texttt{trasta298/keifu} & Terminal interface for navigating and visualizing Git commit graphs & 2.9 & \colorbox{yellow!25}{\strut Medium} \\
\texttt{tree-sitter/}\newline\hspace{1em}\texttt{tree-sitter} & Incremental parsing library for building syntax-aware programming tools & 3.0 & \colorbox{yellow!25}{\strut Medium} \\
\texttt{tstack/lnav} & Log file navigator & 4.7 & \colorbox{red!15}{\strut Hard} \\
\texttt{tukaani-project/xz} & XZ Utils data compression tools and liblzma library & 3.5 & \colorbox{yellow!25}{\strut Medium} \\
\texttt{typst/typst} & Markup-based typesetting system for producing documents and publications & 5.2 & \colorbox{red!15}{\strut Hard} \\
\texttt{unhappychoice/}\newline\hspace{1em}\texttt{gittype} & Terminal typing game that uses source code as typing challenges & 4.6 & \colorbox{red!15}{\strut Hard} \\
\texttt{universal-ctags/}\newline\hspace{1em}\texttt{ctags} & A maintained ctags implementation & 4.2 & \colorbox{red!15}{\strut Hard} \\
\texttt{wfxr/code-minimap} & Renders a scrollable code minimap in the terminal & 1.5 & \colorbox{green!15}{\strut Easy} \\
\texttt{wfxr/csview} & Terminal CSV viewer with column alignment and Unicode support & 1.9 & \colorbox{green!15}{\strut Easy} \\
\texttt{wgunderwood/}\newline\hspace{1em}\texttt{tex-fmt} & Fast LaTeX source code formatter written in Rust & 2.7 & \colorbox{yellow!25}{\strut Medium} \\
\texttt{wintermute-cell/}\newline\hspace{1em}\texttt{ngrrram} & Terminal typing practice tool for learning keyboard layouts & 1.9 & \colorbox{green!15}{\strut Easy} \\
\texttt{xampprocky/tokei} & Counts lines of code, comments, and blanks across programming languages & 3.2 & \colorbox{yellow!25}{\strut Medium} \\
\texttt{xorg62/tty-clock} & Digital clock displayed in the terminal using ncurses & 0.9 & \colorbox{green!15}{\strut Easy} \\
\texttt{y2z/monolith}$^\dagger$ & Saves complete web pages as a single self-contained HTML file & 3.3 & \colorbox{yellow!25}{\strut Medium} \\
\texttt{yaa110/nomino} & Batch file renaming utility with regex and template support & 2.3 & \colorbox{yellow!25}{\strut Medium} \\
\texttt{yassinebridi/serpl} & Terminal interface for interactive search and replace across files & 3.2 & \colorbox{yellow!25}{\strut Medium} \\
\texttt{yoav-lavi/melody} & Language that compiles to regular expressions for improved readability & 3.0 & \colorbox{yellow!25}{\strut Medium} \\
\texttt{ys-l/flamelens} & Terminal-based flamegraph viewer for performance profile analysis & 2.6 & \colorbox{yellow!25}{\strut Medium} \\
\texttt{zevv/duc} & Disk usage analysis tool suite with multiple visualization options & 3.3 & \colorbox{yellow!25}{\strut Medium} \\
\texttt{zk-org/zk} & Command-line tool for managing a plain text Zettelkasten note collection & 3.7 & \colorbox{yellow!25}{\strut Medium} \\
\end{longtable}
}

\end{document}